\documentclass[pra,floatfix,showpacs,amsmath,twocolumn]{revtex4}
\usepackage{amssymb}
\usepackage{graphicx}
\usepackage{graphics}
\usepackage{amsmath}
\usepackage{amsthm}

\newtheorem{theorem}{Theorem}
 \newtheorem{lemma}[theorem]{Lemma}

\newcommand{\bea}{\begin{eqnarray}}
\newcommand{\eea}{\end{eqnarray}}
\def\bi{\begin{itemize}}
\def\ei{\end{itemize}}
\def\bc{\begin{center}}
\def\ec{\end{center}}

\newcommand*{\cP}{\mathcal{P}}

\newcommand*{\cD}{\mathcal{D}}

\def\<{\langle}
\def\>{\rangle}

\def\opone{\leavevmode\hbox{\small1\kern-3.8pt\normalsize1}}

\def\Chi{{\chi}}
\def\ket#1{|#1\rangle}

\newcommand{\one}{\mbox{$1 \hspace{-1.0mm}  {\bf l}$}}
\def\tr{\mathrm{tr}}
\def\ket#1{\left| #1\right>}
\def\bra#1{\left< #1\right|}
\def\ket#1{| #1\rangle}
\def\bra#1{\langle #1 |}

\newcommand{\proj}[1]{\ket{#1}\bra{#1}}

\newcommand*{\bbN}{\mathbb{N}}

\newcommand*{\cE}{\mathcal{E}}

\newcommand*{\eps}{\varepsilon}

\newcommand*{\bn}{\bar{n}}
\newcommand*{\mus}{\mu}
\newcommand*{\mub}{\bar{\mu}}
\newcommand*{\lambdab}{\bar{\lambda}}
\newcommand*{\rhob}{\bar{\rho}}

\newcommand{\cancel}[1]{}

\begin{document}
\title{Security of quantum key distribution protocols using two-way
  classical communication or weak coherent pulses}

\author{Barbara Kraus$^1$, Cyril Branciard$^2$, and Renato Renner$^3$ }

\address{\mbox{} $^1$ Institute for Theoretical Physics, University of Innsbruck, Austria
\\ $^2$ Group of Applied Physics, University of Geneva, 1211 Geneva 4, Switzerland
\\ $^3$ Department of Applied Mathematics and Theoretical Physics, University of Cambridge, Cambridge CB3 OWA, United Kingdom}
\date{\today}

\begin{abstract}
  We apply the techniques introduced in [Kraus et. al., {\em
    Phys.\ Rev.\ Lett.}, {\bf 95}, 080501, 2005] to prove security of
  quantum key distribution (QKD) schemes using two-way classical
  post-processing as well as QKD schemes based on weak coherent pulses
  instead of single-photon pulses. As a result, we obtain improved
  bounds on the secret-key rate of these schemes.
\end{abstract}

 \maketitle

\section{Introduction}

A fundamental problem in cryptography is to enable two distant
parties, traditionally called \emph{Alice} and \emph{Bob}, to
communicate in absolute privacy, even in presence of an eavesdropper,
\emph{Eve}. It is a well known fact that a secret key, i.e., a
randomly chosen bit string held by both Alice and Bob, but unknown to
Eve, is sufficient to perform this task (one-time pad encryption).
Thus, the problem of secret communication reduces to the problem of
distributing a secret key.

Classical key distribution protocols are typically based on unproven
computational assumptions, e.g., that the task of decomposing a large
number into its prime factors is intractable.  In contrast to that,
the security of \emph{quantum key distribution (QKD)} protocols merely
relies on the laws of physics, or, more specifically, quantum
mechanics. This ultimate security is certainly one of the main reasons
why so much theoretical and experimental effort is undertaken towards
the implementation of secure QKD protocols~\cite{GiRi02,IdqMQ}.

Typically \footnote{This description applies to a large class of
QKD
  protocols. There are, however, certain proposals of QKD schemes
  where the encoding is different~\cite{COW,dps}.}, in the first step of a
QKD protocol, Alice chooses a random bit string and encodes each
bit into the state of a quantum system, which she then sends to
Bob (using a quantum channel). Bob applies a certain measurement
on the received quantum system to decode the bit value. In a
second step, called \emph{sifting}, Alice and Bob publicly
exchange some information about the encoding and decoding of each
of the bits which allows them to discard bit pairs which are not
(or only weakly) correlated.

After this sifting process, Alice and Bob hold a pair of classical
correlated bitstrings, in the following called \emph{raw key pair}.
Alice and Bob can determine the quality of the raw key pair by
comparing the values of some randomly chosen bit pairs (using an
authenticated classical communication channel).  This so-called
\emph{parameter estimation} gives an estimate for the \emph{quantum
  bit error rate (QBER)}, i.e., the ratio of positions for which the
values of the bits held by Alice and Bob do not coincide. A
fundamental principle of QKD is that this error rate also imposes a
bound on the amount of information an adversary can have on the raw
key: The smaller the QBER, the more secret key bits can be extracted
from the raw key.  If the QBER is above a certain threshold, then no
secret key can be generated at all, and Alice and Bob have to abort
the protocol~\footnote{In addition to the QBER, further parameters
  estimated by Alice and Bob (e.g., the sifting rate) might be used to
  bound the adversary's information.}.

The purpose of the remaining part of the protocol, called
\emph{classical post-processing}, is to transform the raw key pair
into a pair of identical and secret keys.  In this article, we
consider classical post-processing which consists of the following
three subprotocols: (i)~\emph{local randomization} (also called
\emph{pre-processing}), where Alice randomly flips each of her bits
with some given probability $q$, (ii)~\emph{error correction}, where
Alice and Bob equalize their strings, and (iii)~\emph{privacy
  amplification}, where Alice and Bob apply some compression function
to their bitstring with the aim to reduce Eve's information on the
outcome. Steps (i)--(iii) described above only require (classical)
\emph{one-way communication} from Alice to Bob. However, in practical
implementations, the error correction is sometimes done with two-way
protocols (e.g., the \emph{cascade protocol} \cite{BrSa94}).

In~\cite{KrRe05,ReKr05}, an information-theoretic technique to
analyze QKD protocols of the type described above has been
presented. In contrast to most previously known methods (e.g.,
\cite{ShPr00}), the technique does not require a transformation of
the key distillation protocol into an entanglement purification
scheme, which makes it very general.  It has been applied to prove
the security of various schemes such as the BB84, the six-state,
the B92, and the SARG protocol~\cite{BB84,BeGi99,Be92,ScAc04}
 (see~\cite{KrRe05,ReKr05} for
an analysis of the first three protocols and~\cite{BrGi05} for an
analysis of the latter). In particular, it has been shown that the
local randomization, i.e., step~(i) described above, increases the
bounds on the maximum tolerated QBER by roughly 10--15 \%.

In this paper, we extend the technique of~\cite{KrRe05,ReKr05}
(Section~\ref{sec:oneway}) and apply it to two classes of QKD
protocols which have not been covered in~\cite{KrRe05,ReKr05}. The
first (Section~\ref{sec:twoway}) is the class of so-called
\emph{two-way protocols}. These use an additional subprotocol,
called \emph{advantage distillation}, which is invoked between the
parameter estimation and the classical post-processing step
described above. In contrast to the classical post-processing
considered in \cite{KrRe05,ReKr05}, advantage distillation uses
two-way communication between Alice and Bob.  Second, we study
protocols which use weak coherent pulses instead of single-photon
pulses (Section~\ref{sec:coh}). For both scenarios, we show that local
randomization increases the secret-key rates.

\section{Information-theoretic analysis of QKD schemes} \label{sec:oneway}

In this section we first review the results presented in
~\cite{KrRe05,ReKr05} and then show then show how they can be
generalized. Throughout this paper we use subscripts to indicate the
subsystems on which a state is defined. Alice and Bob's quantum
systems are labelled by $A$ and $B$, respectively. Similarly, the
classical values obtained by measuring their quantum systems are
denoted by $X$ and $Y$, respectively. Typically, we write $\rho_{AB}$,
or $\rho_n$, to denote the state of all the qubits held by Alice and
Bob, whereas $\sigma_{AB}$ is a two-qubit state.  We will often
consider two-qubit Bell-diagonal states, i.e., states that are
diagonal in the Bell basis,
$\ket{\Phi_{ij}}=(\ket{0,0+i}+(-1)^{j}\ket{1,1+i})/\sqrt{2}$.
$P_{\ket{\Phi}}$ denotes the projector onto the state ${\ket{\Phi}}$.
Furthermore, we denote by $h(x)=-x\log_2(x)-(1-x)\log_2(1-x)$ the
binary entropy function.

\subsection{Review of the technique}

The information-theoretic technique proposed
in~\cite{KrRe05,ReKr05} directly applies to a general class of
quantum key distribution protocols using one-way classical
communication. However, it is required that the protocol can be
represented as a so-called \emph{entanglement-based scheme}, as
described below.

Generally, a QKD protocol uses a set of so-called \emph{encoding
  bases}. We consider the special case where each basis $j$ is defined
by two states $\ket{\phi_j^0}$ and $\ket{\phi_j^1}$, which are used to
encode the bit values $0$ and $1$, respectively. In a
\emph{prepare-and-measure} scheme, Alice repeatedly chooses at random
a bit $i$ and a basis $j$, prepares the state $\ket{\phi_j^i}$, and
sends the state to Bob. Bob then measures the state in a randomly
chosen basis $k$.  This measuring process can be seen as some
filtering operation
$B_k=\ket{0}\bra{\phi_{1,k}^\perp}+\ket{1}\bra{\phi_{0,k}^\perp}$,
where $\ket{\phi_{i,k}^\perp}$ is some state orthogonal to
$\ket{\phi_k^i}$, followed by a measurement in the computational
basis.

In an \emph{entanglement-based} view, the above can equivalently be
described as follows: Alice prepares the two-qubit states
$A_j\ket{\Phi_{00}}$, where $\ket{\Phi_{00}}$ denotes the Bell state
$1/\sqrt{2}(\ket{0,0}+\ket{1,1})$ and $A_j$ is an encoding operator
(for details see \cite{KrRe05}) such that
$\bra{i}A_j\ket{\Phi_{00}}=\ket{\phi_j^i}$. She then sends the
second qubit to Bob and prepares Bob's system at a distance by
measuring her system in the computational basis.  Bob's measurement
is described in the same way as in the prepare-and-measure scheme.

Note that, in an experimental realization of a QKD protocol, one might
prefer to implement a prepare-and-measure scheme. However, when
analyzing the security of a protocol, it is usually more convenient to
consider its entanglement-based version.

As an illustration, consider the BB84 protocol, which uses the
$z$-basis and the $x$-basis are used for the encoding. Using the
above notation, we have $\ket{\phi_0^i}=\ket{i_z}$ and
$\ket{\phi_1^i}=\ket{i_x}$, for $i=0,1$. Hence, the operators
applied by Alice are $A_0=\one$ and $A_1=H$, where $H$ denotes the
Hadamard transformation. Because the bases are orthonormal, the same
operators
describe Bob's measurement as well. 


For the following, we assume that Alice and Bob apply a randomly
chosen permutation to rearrange the order of their qubit pairs, in the
following denoted by $\cP_S$, and, additionally, apply to each of the
qubit pairs at random either the identity or the operation
$\sigma_x\otimes\sigma_x$. (Note that the symmetrization operations
commute with the measurement and can therefore be applied to the
classical bit strings). Then, as shown in~\cite{KrRe05}, the state
$\rho_{A B}$ describing the $N$ qubit pairs shared by Alice and Bob
can generally (after the most general attack by Eve, a so-called
\emph{coherent} attack) be considered to be of a simple form, namely
\begin{equation} \label{eq:symform}
\rho_{AB}=\sum_{n_1, \ldots, n_4} \lambda_{n_1,n_2,n_3,n_4} \cP_S
(P_{\ket{\Phi_{00}}}^{\otimes n_1}\otimes
P_{\ket{\Phi_{01}}}^{\otimes
  n_2}\otimes P_{\ket{\Phi_{10}}}^{\otimes n_3}\otimes
P_{\ket{\Phi_{11}}}^{\otimes n_4}) \ .
\end{equation}
The sum runs over all nonnegative $n_1, \ldots, n_4$ such that
$n_1+n_2+n_3+n_4 = N$. The set of possible values of the coefficients
$\lambda_{n_1,n_2,n_3,n_4}$ depends on the specific protocol and the
parameters estimated by Alice and Bob (e.g., the QBER of the raw key).
Furthermore, one can assume without loss of generality that Eve has a
purification of this state, i.e., the situation is fully described by
a pure state $\ket{\Psi}_{ABE}$ such that
$\rho_{AB}=\tr_E(P_{\ket{\Psi}_{ABE}})$.  (However, as we shall see,
dropping this assumption might lead to better estimates of the key
rate.) After this distribution of quantum information Alice and Bob
measure their systems. Thus they are left with classical bit-strings.

Consider now any situation where Alice and Bob have a classical
pair of raw keys $X^n$ and $Y^n$ consisting of $n$ bits whereas
Eve controls a quantum system $E$.  The \emph{secret-key rate},
i.e., the rate at which secret key bits can be generated per bit
of the raw key, for any one-way protocol (with communication from
Alice to Bob), is given by
\begin{equation} \label{rate}
  r= \lim_{\eps \to 0} \lim_{n\rightarrow \infty} \frac{1}{n} \sup_{U^n \leftarrow X^n}
    S_2^{\eps}(U^n E^n) - S_0^{\eps}(E^n)
    - H_0^\eps(U^n|Y^n )
   \ .
\end{equation}
Here, $S_\alpha^{\eps},H_\alpha^{\eps}$ denote the smooth
\emph{R\'enyi entropies} (also called \emph{min-entropy} if $\alpha =
\infty$ and \emph{max-entropy} if $\alpha = 0$)~\cite{ThesisRe}.
Moreover, the supremum runs over all classical values $U^n$ that can
be computed from (the classical value) $X^n$.

For a QKD protocol as described above (where the distributed state is
of the form of Eq.~\eqref{eq:symform}), formula~\eqref{rate} can be
lower bounded by an expression which only involves two-qubit systems.
More precisely~\cite{KrRe05},
\begin{equation} \label{rategen}
  r\geq
  \sup_{\substack{U \leftarrow X }}
    \, \inf_{\sigma_{A B} \in \Gamma_Q}
    S(U | E) - H(U|Y )  \ ,
\end{equation}
where $\Gamma_Q$ is the set of all two-qubit states $\sigma_{A B}$
(after the filtering operation) which can result from a collective
attack~\footnote{A \emph{collective attack} is an attack where the
  adversary treats each signal sent over the channel identically and
  independently of the other signals.} and which are compatible with
the parameters estimated by Alice and Bob (in particular, the QBER).
Here, $S$ and $H$ denote the von Neumann entropy and its classical
counterpart, the Shannon entropy, respectively. Moreover, $X$ and $Y$
denote the classical outcomes of measurements of $\sigma_{A B}$ (on
$A$ and $B$, respectively) in the computational basis, and $E$ is any
system that purifies $\sigma_{A B}$.  Similarly to the above formula,
the supremum runs over all mappings from $X$ to $U$~\footnote{An even
  tighter lower bound is given by $ r\geq \sup_{U \leftarrow X, V
    \leftarrow U}\inf_{\sigma_{A B} \in \Gamma_Q} S(U | V E) - H(U|Y
  V)$. This includes the possibility that Alice sends some additional
  information, $V$ to Bob. However, we are not aware of any protocol
  where this additional step helps (at least not after the sifting
  phase).}.

\subsection{Local randomization}

The local randomization step described above has first been
considered in~\cite{KrRe05,ReKr05} and later been improved in
~\cite{SmReSm06}. In~\cite{RenSmis06}, the local randomization is
nicely explained in the context of entanglement purification.

To get an intuition why the local randomization can help to increase
the secret-key rate, it is useful to describe the process as a quantum
operation (as in~\cite{RenSmis06}).  Let $\sigma_{A B}$ be the state
of a qubit pair held by Alice and Bob and let $\ket{\Psi}_{ABE}$ be a
purification of $\sigma_{AB}$. The state after Alice randomly flips
her bit value $A$ with probability $q$, can be described by
$\ket{\Psi}_{AA'
  BE}=\sqrt{1-q}\ket{\Psi}_{ABE}\ket{0}_{A'}+\sqrt{q}\sigma_x^A
\ket{\Psi}_{ABE}\ket{1}_{A'}$, where $A^\prime$ is an auxiliary system
on Alice's side. The measurement of system $A$ gives the raw key. Note
that $\ket{\Psi}_{AA' BE}$ results from the application of a
\emph{controlled-not} operation on system $AA^\prime$, where system
$A^\prime$ is prepared in the state
$\sqrt{1-q}\ket{0}_{A'}+\sqrt{q}\ket{1}_{A'}$.  The randomization of
Alice thus entangles her system to some auxiliary system (which is not
under Eve's control).  This, in turn, reduces the entanglement between
Alice's relevant system ($A$) and Eve's systems (monogamy of
entanglement), as Eve does not have a purification of the state on the
systems $A$ and $B$, since now she only has the purification of the
state $\rho_{AA^\prime B}$.  Note that Bob's information on $A$ is
also reduces by the randomization process, but---for certain values of
the parameter $q$---he is less penalized than Eve. From this point of
view, it can be easily understood that the local randomization can
help to increase the secret-key rate.

\subsection{Comparison to known bounds}

For protocols based on qubit pairs, where the raw key pair is
obtained by orthogonal measurements of Alice and Bob on some
Bell-diagonal state $\sigma_{A B}=\sum_{i,j} \lambda_{ij}
P_{\Phi_{ij}}$ (e.g., the BB84 or the six-state protocol), it
follows from~\eqref{rategen} that the secret-key rate $r$ (even
without the local randomization) is bounded by \[r \geq
1-S(\sigma_{AB}) \geq 1-h(e_b)-h(e_p) \ .
\] Here, $e_b=\lambda_{10}+\lambda_{11}$ is the QBER and
$e_p=\lambda_{01}+\lambda_{11}$ the \emph{phase error rate}, i.e.,
the probability that Alice and Bob get different bits when measuring
in the $z$ and the $x$-basis, respectively. Because the QBER and the
phase error rate are not changed by applying at random $\sigma_x$ or
$\sigma_z$, which make any state Bell diagonal, the bound
$1-h(e_b)-h(e_p)$ holds for arbitrary states $\sigma_{A B}$.  Note
that the above bound implies any of the lower bounds on the one-way
secret-key rate derived in previous works~\cite{ShPr00,Lo01}.

\subsection{Generalization of the lower bound} \label{improve_bound}

Because we assume above that Eve controls a system that purifies the
state $\rho_{A B}$ held by Alice and Bob, the bound~\eqref{rategen} is
fully determined by $\rho_{A B}$.  However, this assumption on Eve
might overestimate her possibilities, in which case the bound is not
optimal. In the following we drop this assumption to derive better
lower bounds on the secret-key rate.

Suppose that the state distributed in an entanglement-based scheme is
of the form $\cP_S((\cD_{AB} \otimes \one)^{\otimes n}(\rho^0_{A B
  E}))$, where $\cP_S$ again denotes the map that randomly permutes
the order of the qubit pairs, $\cD_{AB}$ is some completely positive
map on two-qubit states, and $\rho^0_{A B E}$ is some tripartite
state. Then, it is an immediate consequence of Lemma~A.4
in~\cite{ReKr05} that the bound~\eqref{rategen} on the secret-key rate
can be generalized to
\begin{equation} \label{rategenmixed}
  r\geq
  \sup_{\substack{U \leftarrow X }}
    \, \inf_{\tilde{\sigma}_{ABE} \in \tilde{\Gamma}_Q}
    S(U | E) - H(U|Y )  \ .
\end{equation}
Here, the infimum ranges over the set $\tilde{\Gamma}_Q$ of all states
$\tilde{\sigma}_{ABE}$ which can result from a collective attack and
are compatible with the parameters estimated by Alice and Bob (e.g.,
the QBER).

We refer to Appendix~\ref{app:SARGimprove} for an application of
this result to improve the analysis of the one-way SARG protocol
for single-photon pulses.

Consider now the general situation where the state describing Alice,
Bob, and Eve's system is the reduced density operator of a state
$\ket{\Psi}_{ABER}=\sum_n \alpha_n \ket{\Psi_n}_{ABE}\ket{n}_R$,
where $\{\ket{n}\}$ forms an orthonormal basis of the Hilbert space
of an auxiliary system $R$, i.e., none of the three parties has the
auxiliary system at their disposal. Starting
from~\eqref{rategenmixed} and using the concavity of the entropy, we
find that the secret-key rate is bounded by
\begin{equation} \label{eq:rateconv}
  r \geq \sup_{U \leftarrow X} \inf_{\tilde{\sigma}_{ABE} \in \tilde{\Gamma}_Q} \Bigl(\sum_{n=0}^\infty |\alpha_n|^2 S(U| E,n) \Bigr)-H(U|Y),
\end{equation}
where $S(U|E, n)=S(UE|n)-S(E|n)$, is the entropy of $U$ conditioned on
$E$ and the event that the measurement of the auxiliary system $R$ in
the basis $\{\ket{n}\}$ yields $n$.

One might also improve the bound using the following observation
which has also been used to derive the bound given in
Eq.~\eqref{rategen}. Let us consider the situation where some
auxiliary system is at Alice' and/or Bob's disposal, but not at
Eve's (this could be for instance some additional qubits). Suppose
that the state shared by $ABE$ and some auxiliary system $R$
(which is not under Eve's control) is given by
$\ket{\Psi}_{ABER}=\sum_n\alpha_n\ket{\Psi_n}_{ABE} \ket{n}_R$,
where $\{\ket{n}\}$ is an orthonormal basis of ${\cal H}_R$, the
Hilbert space corresponding to system $R$. The state
$\ket{\tilde{\Psi}}_{ABER}=\sum_n \alpha_n
U^{AB}_n\ket{\Psi_n}_{ABE} \ket{n}_R$, with $U^{AB}_n$ unitary
operators diagonal in the $z$-basis leads to the same measurement
outcome for any measurement by Alice and Bob in the computational
basis as $\ket{\Psi}_{ABER}$, that is $\ket{k,l}_{AB}\bra{k,l}
\rho_{ABE}\ket{k,l}_{AB}\bra{k,l}=\ket{k,l}_{AB}\bra{k,l}
\tilde{\rho}_{ABE}\ket{k,l}_{AB}\bra{k,l}$, where
$\rho_{ABE}=\tr_R(P_{\ket{\Psi}_{ABER}})$ and
$\tilde{\rho}_{ABE}=\tr_R(P_{\ket{\tilde{\Psi}}_{ABER}})$.
Assuming that Eve has a purification of the state
$\tilde{\rho}_{AB}$ can only provide her with more power compared
to the situation where she has a purification of the state
$\rho_{ABR}$, since this is equivalent to giving her the system
$R$, which she could simply measure, leading to the same result as
before (for details see also \cite{KrRe05}). Thus, we can consider
the situation where Alice and Bob share the state
$\tilde{\rho}_{AB}$ and Eve has a purification of it. This can
only increase Eve's power. We will use this observation in
Appendix B, in order to determine a good lower bound on the
secret-key rate for a QKD protocol using the so-called XOR
process.

\section{QKD protocols with two-way post-processing} \label{sec:twoway}

In the following, we will consider QKD protocols where, before the
post-processing of the raw key as described above, Alice and Bob
additionally invoke a so-called \emph{advantage-distillation}
subprotocol, which requires two-way communication between Alice and
Bob.  The notion of advantage distillation has first been
investigated in the context of classical key
agreement~\cite{Maurer93} and later been generalized to
QKD~\cite{GoLo03,Ch02}.

The advantage distillation protocol we consider here has the following
form: Alice publicly announces to Bob the position of a block of $m$
bits which have all the same value (of course, she does not tell him
which value).  Then Bob tells Alice whether for the given position,
his corresponding bits are all identical as well. If this is the case,
they both continue using the first bit of the block as a new raw-key
bit, otherwise they discard the whole block. We emphasize here that
our analysis below works for any fixed value of the block size $m$
(not only asymptotically for large $m$). This is important for
realistic protocols, where $m$ is usually small (e.g., $m=3$).

To simplify the study of such protocols, we first show that it
suffices to analyze the action of the advantage distillation process
on two-qubit Bell-diagonal states.  More precisely,
Lemma~\ref{lem:prod} below implies that the state $\bar{\rho}_{\bn}$
obtained by applying a block-wise operation $\cE$ (for blocks of size
$m$) to a symmetric state $\rho_n$ (see Eq.~\eqref{eq:symform}) has
virtually the same statistics as if $\cE$ was applied to a state
$\sigma^{\otimes m}$.

\begin{lemma} \label{lem:prod}
  Let $\rho_n$ be a state on $n$ particle pairs of the form
  \[
    \rho_n
  =
    \cP_S (P_{\ket{\Phi_{00}}}^{\otimes n_1}\otimes
  P_{\ket{\Phi_{01}}}^{\otimes n_2}\otimes P_{\ket{\Phi_{10}}}^{\otimes
  n_3}\otimes P_{\ket{\Phi_{11}}}^{\otimes n_4})
  \]
  and let $\sigma$ be a two-qubit Bell-diagonal state with eigenvalues
  $\frac{n_1}{n}, \ldots, \frac{n_4}{n}$. Moreover, let $\cE$ be an
  operation which maps Bell states of blocks of $m$ particle pairs to
  Bell states of one single particle pair. Finally, let
  \[
    \rhob_{\bn}
  =
      \sum_{\bn_1, \ldots, \bn_4} \mub_{\bn_1,\bn_2,\bn_3,\bn_4}
    {\cal P}_S (P_{\ket{\Phi_{00}}}^{\otimes \bn_1}\otimes
  P_{\ket{\Phi_{01}}}^{\otimes \bn_2}\otimes P_{\ket{\Phi_{10}}}^{\otimes
  \bn_3}\otimes P_{\ket{\Phi_{11}}}^{\otimes \bn_4})
  \]
  be the state describing $\bn = \frac{n}{m}$ particle pairs defined by
  $\rhob_{\bn} := \cE^{\otimes \bn}(\rho_n)$ and let $\lambdab_1,
  \ldots, \lambdab_4$ be the eigenvalues of $\bar{\sigma} :=
  \cE(\sigma^{\otimes m})$. Then, for any $\eps \geq 0$,
  \[
    \sum_{(\bn_1, \ldots, \bn_4) \in \mathcal{B}^\eps(\lambdab_1, \ldots,
\lambdab_4)} \mub_{\bn_1,\bn_2,\bn_3,\bn_4}
  \geq
    1 - 2^{-\Theta(\bn \eps^2) + O(\log n)} ,
  \]
  where $\mathcal{B}^\eps(\lambdab_1, \ldots, \lambdab_4)$ denotes the
  set of all tuples $(\bn_1, \ldots, \bn_4)$ such that
  $(\frac{\bn_1}{n}, \ldots, \frac{\bn_4}{n})$ is $\eps$-close to
  $(\lambdab_1, \ldots, \lambdab_4)$ and $\Theta(\bn \eps^2)$ is
  asymptotically the same as $\bn \eps^2$, up to a constant factor.
\end{lemma}

The lemma is a direct consequence of the exponential quantum de
Finetti Theorem~\cite{ThesisRe}.  It states that, for any
$n$-partite quantum state $\rho_n$ which is invariant under
permutations of the subsystems, any part $\rho_m =
\tr_{n-m}(\rho_n)$ consisting of $m$ subsystems is exponentially
(in $n-m$) close to a convex combination of states that virtually
are of the form $\sigma^{\otimes m}$. For completeness, we give a
direct proof of Lemma~\ref{lem:prod} (without referring to de
Finetti's theorem) in Appendix~\ref{app:proof}.

In order to analyze protocols with advantage distillation using
Lemma~\ref{lem:prod}, we use the following quantum mechanical
description of the advantage distillation subprotocol: Alice and Bob
both apply the operation $X^m_{ad}=\ket{0}\bra{0,\ldots,
  0}+\ket{1}\bra{1,\ldots, 1}$ on $m$ qubits. It is straightforward to
check that \bea (X^2_{ad})^{\otimes
2}(\ket{\Phi_{i,j}}\ket{\Phi_{k,l}})=\frac{1}{\sqrt{2}}\delta_{i,k}\ket{\Phi_{i,j+l}},\eea
where the sum $j+l$ of indices is understood to be modulo~$2$.
Hence, applying advantage distillation to $m$ identical
Bell-diagonal qubit-pairs with eigenvalues
$\lambda$~\footnote{$\lambda$ is a vector
  of eigenvalues $\lambda_{ij}$ corresponding to the Bell states
  $\ket{\Phi_{ij}}$.} leads to a Bell-diagonal state with
eigenvalues $\lambda'$ given by
\begin{align}
 \label{addist}
 \lambda_{i,j}^{\prime}&=
 \frac{1}{T}\big[(\lambda_{i,0}+\lambda_{i,1})^m+ (-1)^j
 (\lambda_{i,0}-\lambda_{i,1})^m\big] \, \end{align} where
$T=2[(1-Q)^m+Q^m]$ and where $Q=\lambda_{10}+\lambda_{11}$ is the QBER
before the advantage distillation. The QBER $Q'$ after the advantage
distillation is thus given by
$Q^\prime=\lambda_{10}^{\prime}+\lambda_{11}^{\prime}=\frac{Q^m}{(1-Q)^m+Q^m}$
and $(1-Q)^m+Q^m$ is the probability that the advantage distillation
is successful (i.e., Alice and Bob end up with a new raw key bit). If
Alice and Bob apply, after the advantage distillation the one-way
classical post-processing described above, the lower bound on the
secret-key rate is given by Eq.~\eqref{rategen}, where the eigenvalues
of $\sigma_{AB}$ are given by the $\lambda$`s in~\eqref{addist}
\footnote{Note that assuming that Eve has a purification of the state
  describing Alice' and Bob's system takes into account the fact that
  Eve knows the classical information, about the bits which are
  grouped in the different blocks.}. For instance for the six-state
protocol one obtains a positive key rate for any $QBER< 0.276$ (for
$m\longrightarrow \infty$). Note that for the six-state protocol it
has been shown that the tolerable QBER cannot be larger than $0.276$,
if the first step in the post-processing is advantage distillation
\cite{AcMa}. As mentioned before, the bound on the secret-key rate is
not only valid, for $m\longrightarrow \infty$, but for any value of
the block size on which advantage distillation is applied.

In \cite{Ch02}, Chau considered the secret-key rate obtained when
applying the above described advantage distillation followed by the
XOR transformation, where Alice and Bob locally compute new raw key
bits by taking the XOR of a block of given bits. (For the sake of
completeness we demonstrate in Appendix~\ref{app:xor}, how the XOR
protocol can be included in our analysis.)  Both procedures were
analyzed in the asymptotic limit for infinitely large block sizes. The
result found there is that the six-state protocol tolerates a QBER of
up to $0.276$.  Surprisingly, the same threshold for the QBER can be
obtained, as shown above, by a simpler protocol where the XOR
transformation is replaced by a local randomization on single bits on
Alice's side.  Moreover, the rate of this modified protocol is much
larger than that of Chau's protocol, as local randomization consumes
less bits than the XOR transformation. Note that, as shown recently by
Bae and Acin~\cite{Acin06}, if one omits the local randomization
completely, the protocol still tolerates a QBER of up to $0.276$, but
the secret-key rate for large values of the QBER might be smaller.

\section{Protocols using weak coherent pulses} \label{sec:coh}

\subsection{Preliminaries}

We now consider protocols where Alice does not send single photons to
Bob, but uses weak coherent pulses instead. This scenario is
practically motivated by the fact that, with current technologies, it
is difficult to create single-photons pulses. In fact, many of today's
implementations of QKD rely on weak coherent pulses.

We start with a description of a prepare-and-measure scheme and then
translate it to an equivalent entanglement based scheme, for which we
will prove security.

In the prepare-and-measure scheme, Alice encodes the bit values into
phase randomized coherent states~\footnote{We do not consider the
  situation where Alice also sends a strong reference pulse to Bob. In
  this case, the state Alice would send is of the form
  $\ket{\psi}=\sum_{n \geq 0} \sqrt{e^{-\mu}
    \mu^n/n!}\ket{n}\ket{N-n}$, where $\ket{n}$ denotes the state of
  $n$ photons in a certain mode. Here we consider the situation where
  she sends only the first of these two systems to Bob.}.  More
precisely, she randomly chooses a basis $j$ and encodes the bit value
$k$ into the state $\rho_{j}^k=\sum_{n\geq 0} p_n
\proj{\phi_j^k}^{\otimes n}$, where $\proj{\phi_j^k}^{\otimes 0}$,
denotes the vacuum for any value of $j$ and $k$ and $p_n=e^{-\mu}
\mu^n/n!$, with $\mu$ the mean photon number (for a Poissonian
source~\footnote{Similarly, one could consider any other distribution
  instead of the Poissonian distribution.}).

The description of Bob's measurement depends on the experimental
setup. We focus on the situation where Bob's detectors do not
distinguish between the cases where they receive one or more than
one photons, since with current technology, it is difficult to
count the number of photons. The POVM describing the photon
detector is thus given by the operators $\{D_0^\dagger
D_0,D_1^\dagger D_1\}$, with $D_0=\sum_{n\geq 0}
\sqrt{p_{n.d.}(n)}P_{\ket{n}}$ and $D_1=\sum_{n\geq 0}
\sqrt{1-p_{n.d.}(n)}P_{\ket{n}}$, where $p_{n.d.}(n)$ is the
probability of not detecting any photon in case $n$ photons
arrived at the detector. This probability is given by
$p_{n.d.}(n)=(1-p_d)(1-\eta)^n$, where $p_d$ is the probability of
a dark count, and $\eta$ is the detection efficiency, i. e.
overall transmission factor. The POVM element $D_0$ corresponds to
the case where no photon is detected, whereas $D_1$ corresponds to
the detection of one or more photons. In the prepare-and-measure
scheme Bob would randomly choose a basis $j$ and measure the
arriving photons in that basis.

In the following, we consider the so-called \emph{untrusted-device
  scenario}, where it is assumed that Eve exchanges Bob's detectors
with perfect ones (having perfect efficiency and no dark counts) and
introduces all errors herself~\footnote{We still make the so-called
  \emph{fair-sampling assumption}, which means that the errors are
  independent of the measurement bases chosen by Bob.}.  Clearly,
security under this assumption implies security in a situation where
Eve might not be able to corrupt Bob's detectors.  Additionally, we
assume that Bob's detector is constructed in such a way that, whenever
a pulse consisting of more than one photon arrives, then the detector
output corresponds to the measurement of one of the photons in the
pulse chosen at random~\footnote{This means that, whenever Bob
  measures a double-click, he has to replace it by a random single
  click.}.

In the described scenario, we can without loss of generality assume
that Eve only sends single photons to Bob. This follows directly from
the fact that the situation obtained by sending a multi-photon pulse
is the same as if Eve randomly selected one photon from the pulse and
sent this single photon to Bob.  Bob's measurement can therefore
simply be described by the operators
$B_j=\ket{0}\bra{\phi_{1,j}^\perp}+\ket{1}\bra{\phi_{0,j}^\perp}$ as
defined previously.

Alice and Bob can estimate the following parameters related to their
raw key: (i)~the total sifting rate $R_{\mu}:=\sum_{n} R_n$, for $R_n
:= p_n Y_n$ where $Y_n$ is the probability for Bob to find a
conclusive result in case Alice sent $n$ photons; (ii)~the average
QBER $Q_{\mu}=\sum_{n} \frac{R_n}{R_\mu} Q_n$, where $Q_n$ denotes the
QBER for the pairs where Alice sent an $n$-photon pulse. These two
parameters will determine the amount of key that can be extracted from
the particular raw key.

We use similar techniques as in \cite{KrRe05,ReKr05} to describe the
same protocol in the entanglement-based scheme. The states prepared
by Alice are \bea \label{weakstate}\ket{\Psi_j}_{ABR_1}=\sum_{n\geq
  0}\sqrt{p_n}\ket{\Psi_j^n}_{AB}\ket{n}_{R_1},\eea where
$\ket{\Psi_j^n}_{AB}=1/\sqrt{2}(\ket{0}_A\ket{\phi_j^0}_B^{\otimes
  n}+\ket{1}_A\ket{\phi_j^1}_B^{\otimes n})$.  Here,
we have introduced an auxiliary system $R_1$ containing the photon
number (which is neither controlled by Alice nor Bob). If Alice
measures her qubit in the computational basis and receives outcome
$k$, the state Bob is left with in the noiseless case (without
interaction of Eve) is $\rho_B=2\tr_{R_1}( P_{\langle k
  \ket{\Psi_j}_{ABR_1}})= \sum_{n\geq 0}p_n
P_{\ket{\phi_j^k}^{\otimes n}}$, which corresponds to the coherent
state (with randomized phase) sent by Alice in the prepare-and
measure scheme \footnote{The factor $2$ after the first equality
sign is due
  to the renormalization (each of Alice's outcome $k$ has probability
  $\frac{1}{2}$).}. The operation on Bob's side is given by the
operators $B_j$, as described above.

The state describing the situation after Bob's operation is given by
\[
\ket{\chi}_{ABE R_1 R_2}=\sum_j B_j
U_{EB}(\ket{\Psi_j}_{ABR_1})\ket{j}_{R_2} \ ,
\]
where $j$ corresponds to the basis chosen by Alice and $U_{E B}$ is a
unitary describing the attack of Eve. Note that this state is not
necessarily normalized, but its weight $\tr(\proj{\chi})$ corresponds
to the sifting rate.

Restricted to Alice and Bob's systems, $\ket{\chi}_{ABE R_1 R_2}$ is a
two-qubit state. We can thus apply the techniques presented in
Section~\ref{sec:oneway} to analyze the security of the protocol. More
precisely, we need to evaluate the r.h.s.\ of~\eqref{eq:rateconv} to
get a lower bound on the secret-key rate.  First we do not take the
local randomization into account, i.e., we choose $U=X$. The case
including local randomization will be treated in the next subsection.
We thus obtain, for the key rate \bea \label{rateweak} r\geq
\inf_{\sigma\in \Gamma_{R_\mu,Q_\mu}} \sum_{n=0}^\infty R_n S(X|E,n)-
R_\mu S(X|Y). \eea The set $\Gamma_{R_\mu,Q_\mu}$ contains all states
which can result from a collective attack by Eve and are compatible
with the average sifting rate $R_\mu$ and the QBER $Q_\mu$, as
estimated by Alice and Bob.

Because the (conditional) entropy of a classical variable cannot be
negative, the r.h.s.\ of~\eqref{rateweak} can be lower bounded by
restricting to any of the terms in the sum over $n$.  Note that,
in~\eqref{rateweak}, the average over $n$ is only taken over the term
for the entropy conditioned on Eve's system, but not on the term for
the entropy conditioned on Bob's system. This is because Eve might be
able to measure the photon number, whereas this is not the case for
Bob.

\subsection{Protocols with local randomization}

So far we did not consider the possibility for Alice to apply some
local randomization on her classical bits. The randomization can
easily be included in the analysis: if the randomization is acting on
single bits, $U \leftarrow X$ (bit flip with probability $q$),
\eqref{rateweak} simply writes \bea \label{rateweak_preproc0} r\geq
\inf_{\sigma\in \Gamma_{R_\mu,Q_\mu}} \sum_{n= 0}^\infty R_n S(U|E,n)-
R_\mu S(U|Y).  \eea Bob's uncertainty is now given by $S(U|Y) =
h(Q_{\mu}^q)$, where $Q_{\mu}^q = (1-q)Q_{\mu} + q(1-Q_{\mu})$.  Since
$R_{\mu} = \sum_n R_n$, \eqref{rateweak_preproc0} can also be written
as \bea
\label{rateweak_preproc} r\geq \inf_{\sigma\in \Gamma_{R_\mu,Q_\mu}}
\sum_{n= 0}^\infty R_n \big[ S(U|E,n) - h(q) \big] \\ \nonumber -
R_\mu \big[ h(Q_{\mu}^q) - h(q) \big]. \eea Note that, for any $n \geq
0$, the term $S(U|E,n)$ on the r.h.s.\ of this inequality can be
bounded by $S(U|E,n) \geq S(U|X) = h(q)$ (since $U$ is only computed
from $X$), and therefore the r.h.s.\ of~\eqref{rateweak_preproc} can
again be lower bounded by restricting the sum to any of its terms.

As we will see, the local randomization allows us to get better lower
bounds for the secret-key rate as well as better lower bounds for the
maximum distance for which the rate is positive.

\subsection{Examples: the BB84 and the SARG protocols}

\label{weak_no_decoy}

Using the results above, in particular~\eqref{rateweak}, we now
compute the lower bound on the secret-key rate of the BB84 as well as
the SARG protocols. In Section \ref{relatedwork} we compare the
results we derive here with previous results, in particular with the
ones presented in~\cite{LoMaChen} and~\cite{FuTaLo}.

In contrast to the single-photon case, where the lower bound on the
secret-key rate was a function of the QBER, we are aiming here for a
lower bound that depends on the only two measurable quantities $R_\mu$
(the total sifting rate) and $Q_\mu$ (the total QBER).  For
simplicity, we will in the following not explicitly include the local
randomization, except in the final results (see
Figures~\ref{fig_no_decoy_V_1} and~\ref{fig_no_decoy_V_095}). We
remind the reader that, in order to include the local randomization,
\eqref{rateweak} simply has to be replaced
by~\eqref{rateweak_preproc}.

Our computation of the bound given by~\eqref{rateweak} is subdivided
into two steps: First, for any $n \geq 0$ and for any $Q_n$, we
compute $S_n(Q_n) := \inf_{\sigma_n \in \Gamma_{Q_n}} S(X|E,n)$, where
$\Gamma_{Q_n}$ is the set of all states $\sigma_n$ which can result
from a collective attack on a $n$-photon pulse causing a QBER of
$Q_n$.  In a second step, we compute the infimum \bea \inf_{ \{R_n,
  Q_n\} \in \widetilde{\Gamma}_{R_{\mu},Q_{\mu}}} \sum_{n = 0}^\infty
R_n S_n(Q_n) \eea where $\widetilde{\Gamma}_{R_{\mu},Q_{\mu}}$
denotes the set of all parameters $\{R_n, Q_n\}$ which are
compatible with $R_{\mu}$ and $Q_{\mu}$. All the technical details
can be found in Appendix ~\ref{app_comput_LB_wcp}.

\subsubsection{BB84}

For the BB84 protocol, it is easy to verify that for any pulse
consisting of $n \geq 2$ photons, Eve has full information on Alice's
measurement outcome $X$, i.e., $\inf_{\sigma_n \in \Gamma_{Q_n}}
S(X|E,n) = 0$ $\forall n\geq 2$.  The lower bound is thus given
by~\footnote{In the untrusted-device scenario, with Bob's detector
  replaced by the eavesdropper, Eve should not send any photon to Bob
  when she receives an empty pulse from Alice, and therefore $R_0 =
  0$. In the trusted-device scenario however, the dark counts could
  contribute to the key with a positive term $R_0$. For a similar
  observation, see~\cite{Lo05}.} \bea r \geq \inf_{\{R_1, Q_1\} \in
  \widetilde{\Gamma}_{R_{\mu},Q_{\mu}}} \ R_1 S_1^{\mathrm{BB84}}(Q_1)
\ - R_{\mu} h(Q_{\mu}) \label{LB_BB84} \eea where
$S_1^{\mathrm{BB84}}(Q_1) := 1 - h(Q_1)$ (see
Appendix~\ref{app_comput_LB_wcp} or \cite{KrRe05,ReKr05}).

As shown in Appendix~\ref{app_comput_LB_wcp}, the conditions in the
untrusted-device scenario for $R_1$ and $Q_1$ to be compatible with
$R_{\mu}$ and $Q_{\mu}$ are the following:
\bea \begin{array}{rcl} R_1 & \leq & \frac{1}{2} p_1 \\
R_1 & \geq & R_{\mu} - \frac{1}{2} \sum_{n \geq 2} p_n\\
R_1 Q_1 & \leq & R_{\mu} Q_{\mu}.
\end{array} \label{constr_BB84} \eea
Let $R_1^{\min} = R_{\mu} - \frac{1}{2} \sum_{n \geq 2} p_n$. If
$R_1^{\min} \leq 0$, then $R_1$ can be set equal to zero, and the
lower bound on $r$ is negative, i.e., Alice and Bob have to abort
the protocol. If $R_1^{\min} > 0$, let $Q_1^{\max} = \min(R_{\mu}
Q_{\mu} / R_1^{\min}, \frac{1}{2})$. Due to the decreasing of
$S_1^{\mathrm{BB84}}(Q_1)$ for $Q_1 \leq 1/2$, we then get \bea r
\geq R_1^{\min} (1-h(Q_1^{\max})) \ - R_{\mu} h(Q_{\mu}).
\label{LB_BB84_bis}\eea

Note that this bound has first been derived in~\cite{GLLP} using a
different technique. This bound can be interpreted as follows: For an
optimal attack, Eve should make $R_1$ as small as possible (i.e.,
block as many single-photon pulses as possible) and, at the same time,
make $Q_1$ as large as possible (i.e., introduce as many errors as
possible on the single-photon pulses that she forwards, which reduces
her uncertainty on Alice's system as much as possible).

\begin{center}
\begin{figure}
  \includegraphics[width=8cm]{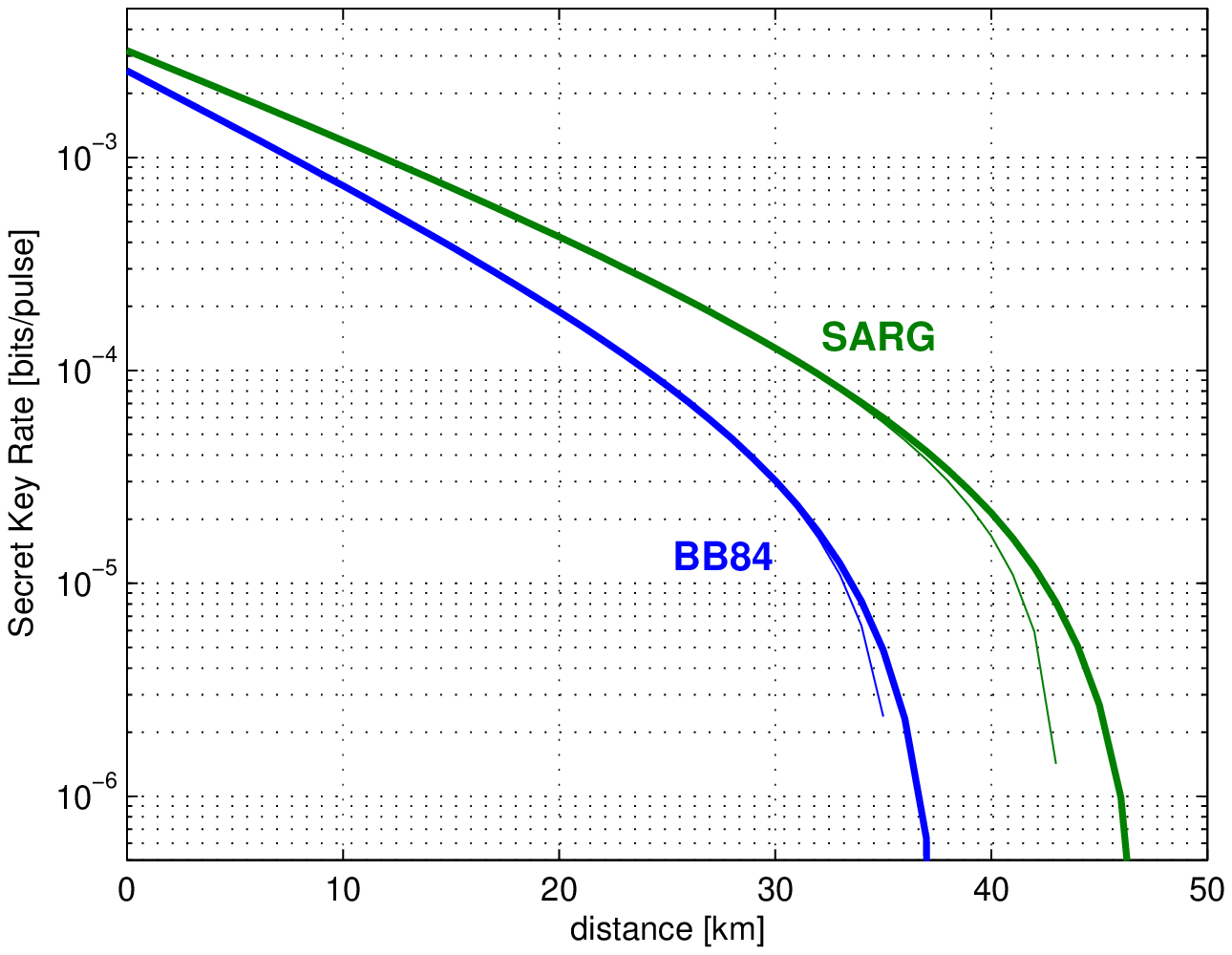} \vspace{2mm}
  \includegraphics[width=8cm]{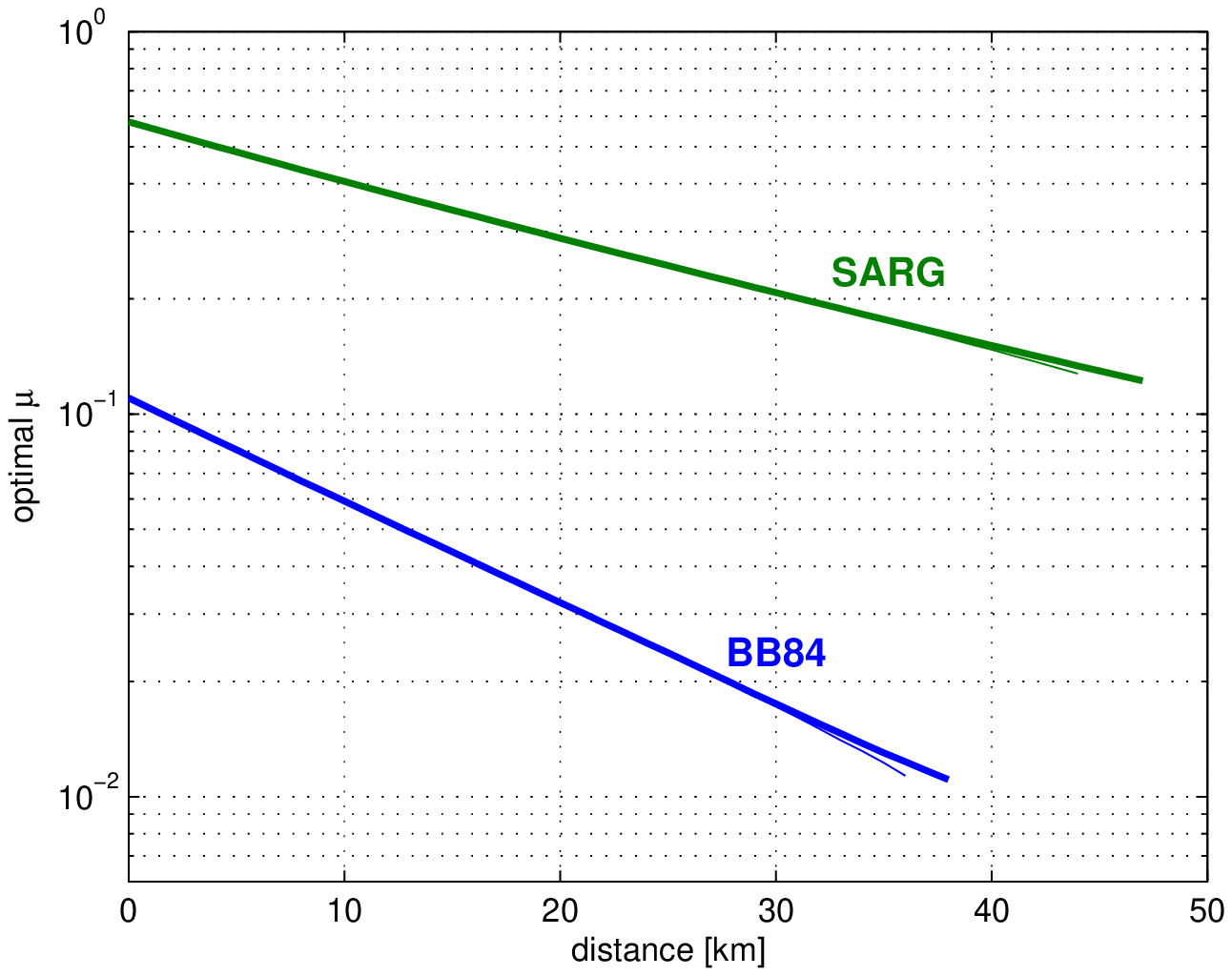}
\caption{Lower bound on the secret-key rate per pulse and optimal
  $\mu$ for Poissonian sources as a function of the distance, for the
  BB84 and SARG protocols, when Alice and Bob share a quantum channel
  with perfect visibility $V = 1$. The other experimental parameters
  are $\alpha=0.25$ dB/km, $\eta_{det}=0.1$ and $p_d=10^{-5}$. The
  thick lines are the results we obtain when Alice performs an optimal
  bit-wise local randomization; the thin lines are the same, without
  randomization ($q=0$).} \label{fig_no_decoy_V_1}
\end{figure}
\end{center}

\begin{center}
\begin{figure}
\includegraphics[width=8cm]{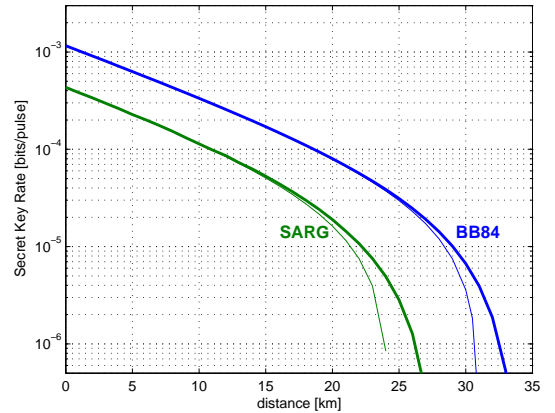}
\caption{Same plot as in Fig. \ref{fig_no_decoy_V_1} (top), but for
a quantum channel with non-perfect visibility, $V = 0.95$.}
\label{fig_no_decoy_V_095}
\end{figure}
\end{center}

To get an idea of how good this bound is, we evaluate the rate for
the situation where there is no Eve present, instead, the errors
are introduced due to a realistic channel. The channel we consider
is a lossy depolarizing channel with visibility $V$ (or fidelity
$F = \frac{1+V}{2}$ and disturbance $D = \frac{1-V}{2}$), and a
transmission factor $t = 10^{-\frac{\alpha \ell}{10}}$ at distance
$\ell$ ($\alpha$ is the attenuation coefficient). Furthermore, we
consider the situation where Bob's detectors have an efficiency
$\eta_{det}$ and a probability of dark counts $p_d$. An explicit
calculation (see Appendix~\ref{app_expected_rates}) shows that
under these assumptions, the rates that Alice and Bob would get
are
\[\begin{array}{lll}
R_{\mu} & = & \frac{1}{2} \big[ 1 - \bar{p}_d^{\ 2} \ e^{-\mu \eta} \big] \\
R_{\mu} Q_{\mu} & = &  \frac{1}{4} \big[ 1 + \bar{p}_d e^{-\mu F
\eta} - \bar{p}_d e^{-\mu D \eta} - \bar{p}_d^{\ 2} \ e^{-\mu \eta}
\big],
\end{array}
\]
where $\eta = t \eta_{det}$, $\bar{p}_d = 1 - p_d$. When we insert
these values in~\eqref{LB_BB84_bis} for experimentally reasonable
values of $\alpha$, $p_d$ and $\eta_{det}$, and optimize for different
distances over the mean photon number $\mu$ (which Alice is free to
choose), we get the results illustrated in Fig.~\ref{fig_no_decoy_V_1}
(for $V = 1$) and Fig.~\ref{fig_no_decoy_V_095} (for $V = 0.95$). We
find that the optimal $\mu$ is proportional to the transmission factor
$t$, and our bound on the secret-key rate is proportional to $t^2$ (at
least for short distances, i.e., in the regime where dark counts
are not dominant); this was already observed in \cite{ILM,GLLP}.

\subsubsection{SARG}

A major difference between the SARG protocol and the BB84 protocols is
that Eve cannot get full information on Alice's value even if the
pulse contains two photons. In order to take this into account, we
include the contribution of the two-photon components in our formula
for the secret-key rate, i.e. we compute \footnote{In the SARG
  protocol, the pulses with three or more photons could still give a
  small contribution to the key (Eve doesn't have full information on
  these). For simplicity we limit ourselves to the 1- and 2- photon
  contributions.}: \bea
\begin{array}{ll} r \geq & \inf_{\{R_1, Q_1, R_2, Q_2\}} \ R_1
S_1^{\mathrm{SARG}}(Q_1) + R_2 S_2^{\mathrm{SARG}}(Q_2) \\ &
\qquad \qquad \qquad \qquad \qquad - R_{\mu} h(Q_{\mu}).
\label{LB_SARG}\end{array}\eea In Appendix~\ref{app_comput_LB_wcp}
we describe how to compute $S_1^{\mathrm{SARG}}(Q_1)$ and
$S_2^{\mathrm{SARG}}(Q_2)$ (see also Appendix C and
\cite{FuTaLo}), and we derive the following conditions for $R_1$,
$Q_1$, $R_2$ and $Q_2$ to be compatible with $R_{\mu}$ and
$Q_{\mu}$:

\bea \begin{array}{rcl} R_1 (1 - Q_1) & \leq & \frac{1}{4} p_1 \\
R_2 (1 - Q_2) & \leq & \frac{1}{4} p_2 \\
R_1 (1 - Q_1) + R_2 (1 - Q_2) & \geq & R_{\mu} (1 - Q_{\mu}) \\ & & \  - \frac{1}{4} \sum_{n \geq 3} p_n\\
R_1 Q_1 + R_2 Q_2 & \leq & R_{\mu} Q_{\mu}.
\end{array} \label{constr_sarg} \eea

If $R_{\mu} (1 - Q_{\mu}) - \frac{1}{4} \sum_{n \geq 3} p_n > 0$,
one can see in \eqref{LB_SARG} that Eve's optimal choice is to set
$R_1$ and $R_2$ as small as possible, and $Q_1$ and $Q_2$ as large
as possible ($S_1^{SARG}(Q_1)$ and $S_1^{SARG}(Q_2)$ are
decreasing): she should therefore set the equality in the third
constraint.

However, contrary to BB84, we have not been able to give a simpler
analytical expression for the infimum in~\eqref{LB_SARG}; we therefore
resort to numerical computations.

Again, in order to estimate the previous bound in a practical
implementation of the protocol, we compute the typical values of the
parameters $R_{\mu}$ and $Q_{\mu}$ when Alice and Bob use a Poisson
source and a lossy depolarizing channel (see
Appendix~\ref{app_expected_rates}):
\[\begin{array}{lll}
R_{\mu} & = & \frac{1}{2} \big[ 1 - \bar{p}_d^{\ 2} \ e^{-\mu \eta}
+ \frac{\bar{p}_d}{2}
e^{-\mu F \eta} - \frac{\bar{p}_d}{2} e^{-\mu D \eta} \big] \\
R_{\mu} Q_{\mu} & = & \frac{1}{4} \big[ 1 - \bar{p}_d^{\ 2} \
e^{-\mu \eta} + \bar{p}_d e^{-\mu F \eta} - \bar{p}_d e^{-\mu D
\eta} \big].
\end{array}
\]

Similarly to the BB84 protocol, inserting these values in
Eq.~\eqref{LB_SARG}, and optimizing for different distances over the
mean photon number $\mu$, provides the results illustrated in
Figures~\ref{fig_no_decoy_V_1} and~\ref{fig_no_decoy_V_095}.

For $V = 1$, we find an optimal $\mu$ proportional to $t^{1/2}$,
and therefore our bound on the secret-key rate scales like
$t^{3/2}$ (see also \cite{Koashi}), which is more efficient than
for BB84 (where we had $r \propto t^2$). For $V = 0.95$ however,
we find that the SARG protocol is less efficient than the BB84,
and our lower bound for the secret-key rate of SARG also scales
like $t^2$, the same as for BB84. However, it should be noted that
we determine here only lower bounds on the rates.

\subsection{Decoy states}

\label{weak_decoy}

The relevant set $\Gamma_{R_{\mu},Q_{\mu}}$ in~\eqref{rateweak} over
which the infimum has to be taken to obtain the lower bound on the
secret-key rate is quite big, since Alice and Bob can only estimate
the total sifting and total error rate.  They do neither have a good
estimation of the error rates, $Q_n$ nor of the corresponding yields,
$Y_n$. Hwang and Lo {\it et. al.} pointed out a method to improve the
lower bound on the secret-key rate by making some additional
measurements (\cite{ref_decoy,LoMaChen} see also \cite{Wang05}). The
idea of the so-called \emph{decoy} states is to change the intensity
of the pulses sent by Alice in order to be able to estimate more
quantities. This allows them to deduce more information about the
possible attack of an eavesdropper (like the estimate of the QBER
does). For practical purpose one assumes that Alice is always sending
weak coherent pulses, varying only the mean photon number. We will
show here how this particular idea can be included in our analysis.

Let us first of all consider the case where Alice uses two different
intensities, i.e., one with mean photon number $\mu_0$ (we call it
signal pulse in the following) and the other (decoy pulse) with mean
photon number $\mu_1$. Using more decoy states is a straightforward
generalization of this case. We describe the states sent by Alice by
$\ket{\psi}_{ABR_1R_2}=\ket{\psi_s}_{ABR_1}\ket{0}_{R_2}+\ket{\psi_c}_{ABR_1}\ket{1}_{R_2}$,
where $\ket{\psi_s}_{ABR_1}$ ($\ket{\psi_c}_{ABR_1}$) denotes the
(unnormalized) signal (decoy) pulse (see Eq.~\eqref{weakstate}).
System $R_2$ is again some auxiliary system, introduced to keep
track of the signal and decoy pulses. In this case this system is in
Alice's hands, as she chooses the intensity of the signals.  Since
Alice is going to measure the auxiliary system $R_2$ in the
computational basis, we can consider the state $\sigma=p_s
\sigma_s\otimes P_{\ket{0}_{R_2}}+(1-p_s)\sigma_c\otimes
P_{\ket{1}_{R_2}}$, where $\sigma_s$ ($\sigma_c$) are Alice and
Bob's signal (decoy) systems after Eve's intervention, respectively.
Bob's measurement is described in the same way as before. Again,
Alice and Bob can only measure the total sifting rate
$R_{\mu}=\sum_{n} R_n=\sum_{n} p_n Y_n$ and estimate the total error
rate $Q_{\mu}=\sum_{n}R_n Q_n / R_{\mu} =\sum_{n}p_n Y_n Q_n /
R_{\mu}$. However, now they are in the position to obtain more
information about their qubit pairs, as they are capable of
measuring these quantities for different values of $\mu$ (recall
$p_n=e^{-\mu} \mu^n/n!$), i.e. they can measure the values
$R_{\mu_0},Q_{\mu_0}$ and $R_{\mu_1},Q_{\mu_1}$. We can again
use~\eqref{rateweak} to compute a lower bound on the secret-key
rate. In this case, the infimum is taken over the set
$\Gamma_{\{R_{\mu_i},Q_{\mu_i}\}_i}$ of all Bell-diagonal two-qubit
states of the form $p_s \sigma_s + p_c \sigma_c$, with $\sigma_s$
($\sigma_c$) denoting the Bell-diagonal states corresponding to the
signal (decoy) bits, which are compatible with all estimated total
sifting rates $R_{\mu_i}$ and total error rates $Q_{\mu_i}$.

\begin{center}
\begin{figure}
\includegraphics[width=8cm]{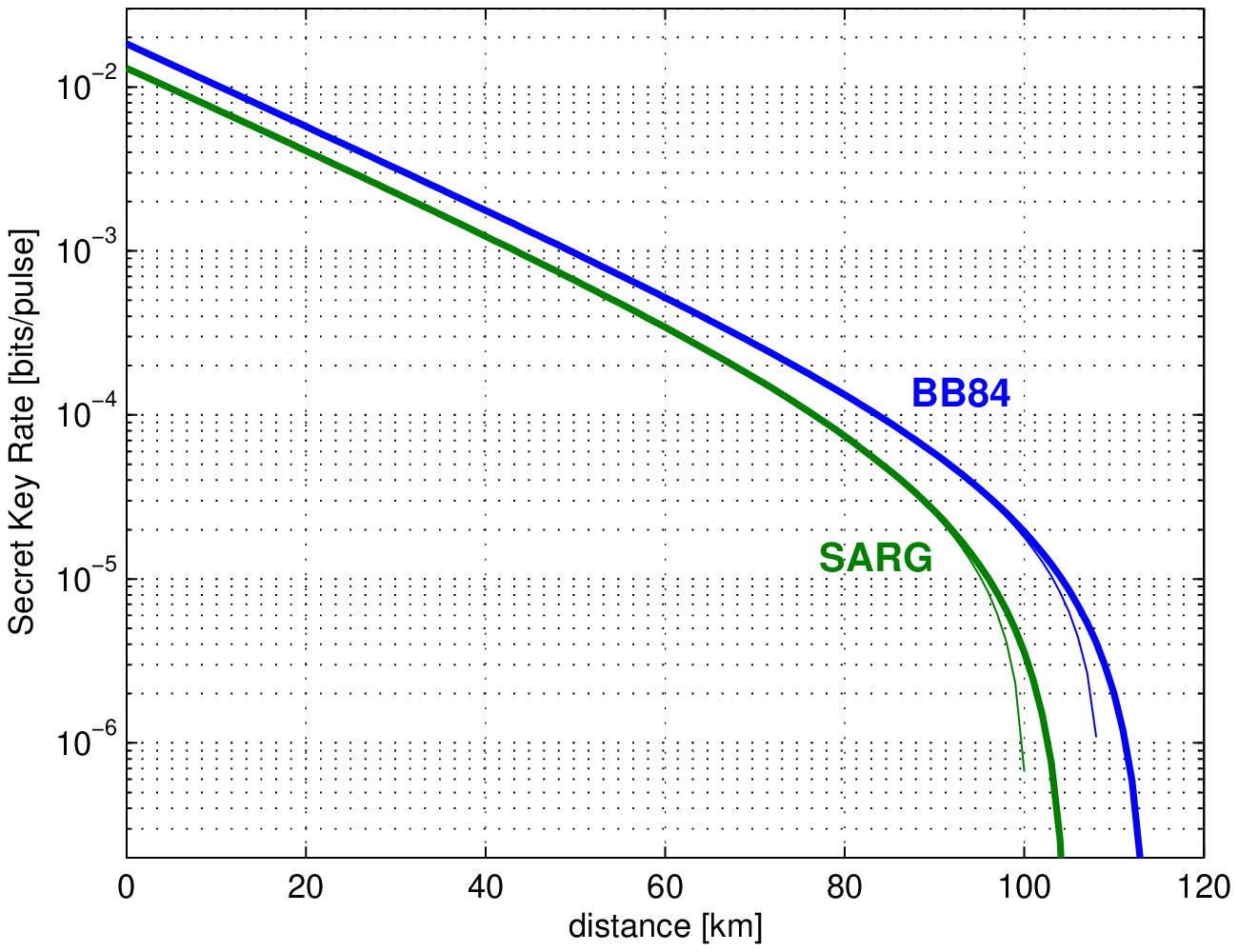}
\vspace{2mm}
\includegraphics[width=8cm]{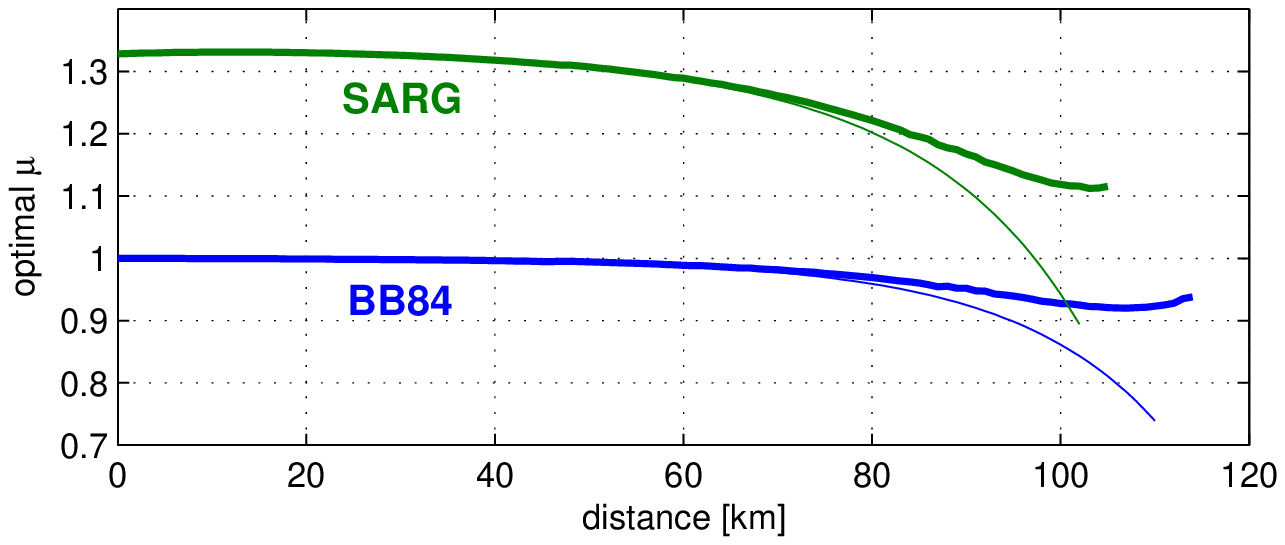}
\caption{Lower bound on the secret-key rate per pulse and optimal
  $\mu$ for Poissonian sources as a function of the distance, for the
  BB84 and SARG protocols using decoy states, when Alice and Bob share
  a quantum channel with perfect visibility $V = 1$. The other
  parameters are the same as in Fig.~\ref{fig_no_decoy_V_1}. The thick
  lines are the results we obtain when Alice performs an optimal
  bit-wise local randomization; the thin lines correspond to the
  protocol without randomization ($q=0$).}
\label{fig_decoy_V_1}
\end{figure}
\end{center}

\begin{center}
\begin{figure}
\includegraphics[width=8cm]{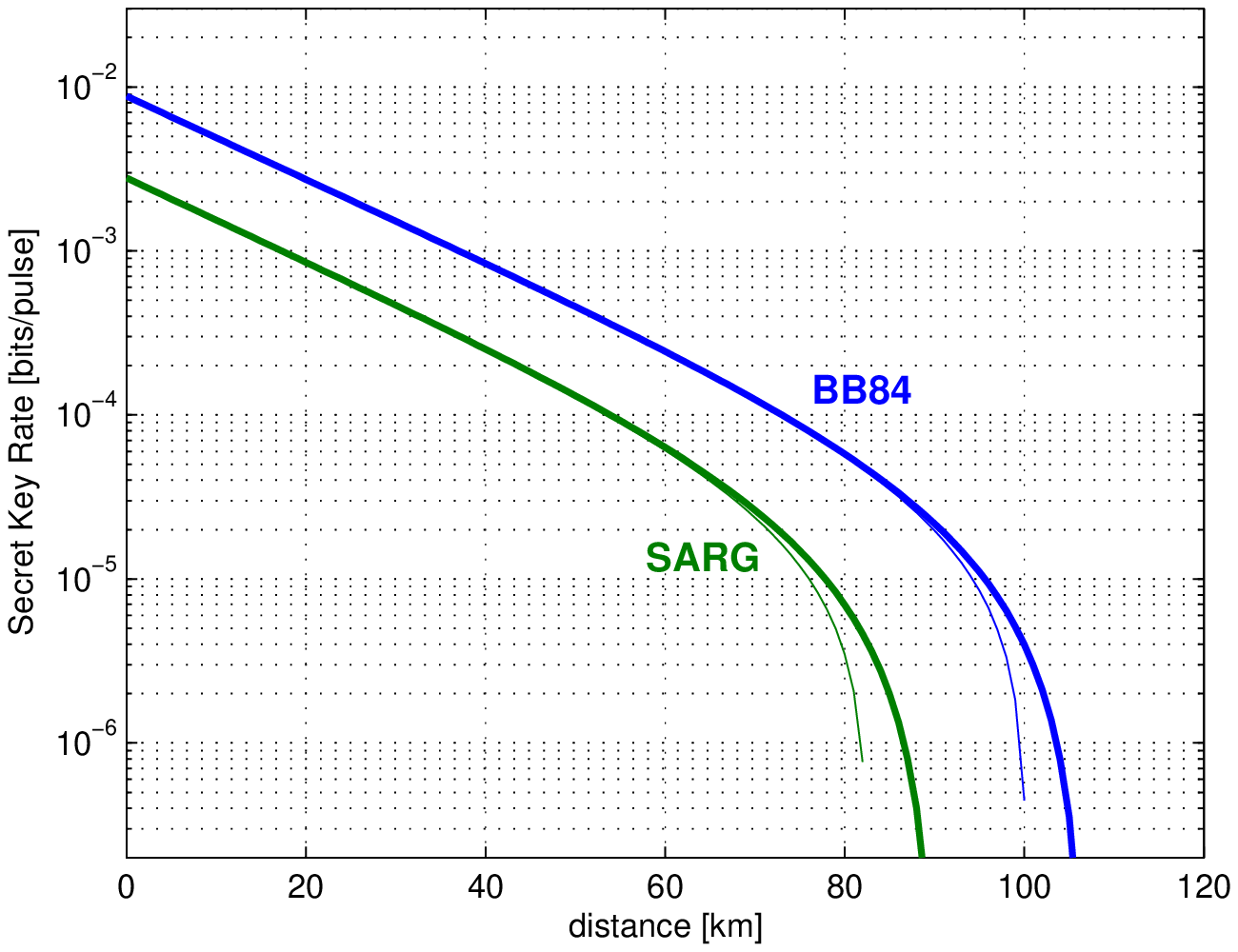}
\vspace{2mm}
\includegraphics[width=8cm]{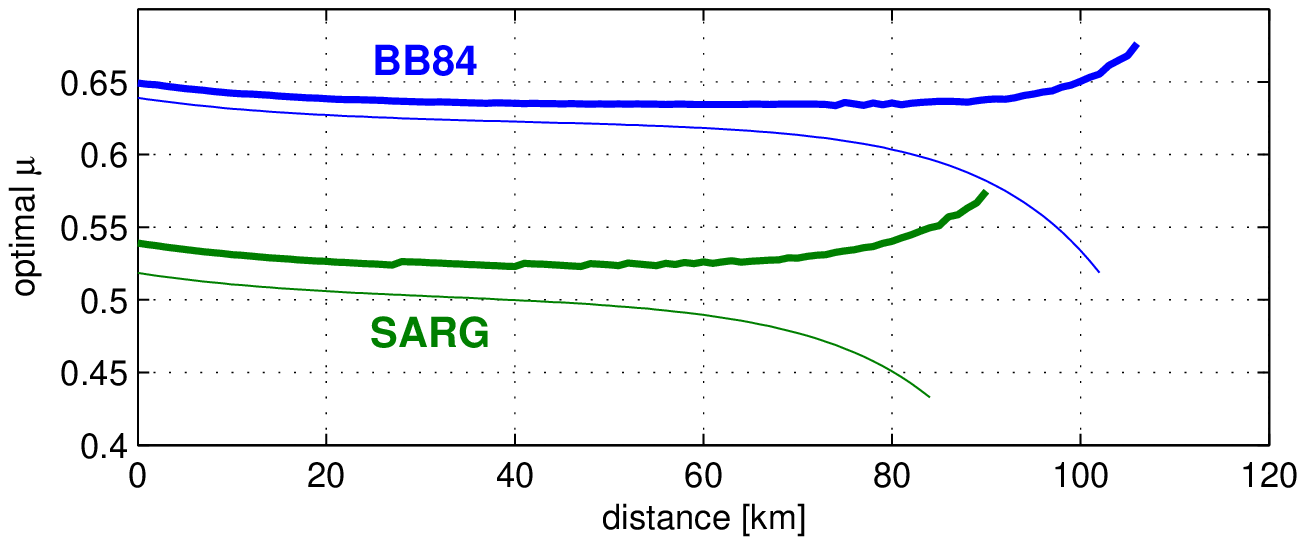}
\caption{Same plots as in Fig. \ref{fig_decoy_V_1}, but for a
quantum channel with non-perfect visibility, $V = 0.95$.}
\label{fig_decoy_V_095}
\end{figure}
\end{center}

Let us now consider the case where Alice uses many different
intensities for her decoy states. Due to the definition of $R_\mu$
it is clear that, by varying $\mu$, one can obtain information about
the quantities $Y_n$. Knowing $Y_n$ and $\{Q_\mu\}$ one can then
determine $Q_n$.  Note that in order to determine $Y_n$ and $Q_n$
one needs infinitely many decoy intensities; however, already a
small number of such decoy intensities suffices to restrict the
values of $Y_n$ and $Q_n$ (see for instance \cite{Wang05}). The
results of the analysis above are illustrated in
Fig.~\ref{fig_decoy_V_1} and~\ref{fig_decoy_V_095}. In order to
evaluate the lower bounds we consider the situation where Alice and
Bob share a lossy depolarizing channel with visibilities $V=1$,
$V=0.95$ respectively.

\subsection{Related work}
\label{relatedwork}

In \cite{FuTaLo}, a similar comparison between the BB84 and SARG
protocols has been done, and lower bounds on the secret-key rates were
computed. For BB84, our results are very similar to those
of~\cite{FuTaLo} (see also \cite{LoMaChen}), but we could slightly
increase the rates and the limiting distances with using the local
randomization process~\footnote{The difference for instance in the
  plots for BB84 in Fig.~9 of \cite{FuTaLo} and our
  figures~\ref{fig_no_decoy_V_1} and \ref{fig_decoy_V_1} essentially
  come from a different definition of the dark count probabilities :
  in \cite{FuTaLo}, $p_{\mathrm{dark}}$ is the probability that one of
  the two detectors has a dark count, while here $p_d$ is the
  probability for each detector to have a dark count (therefore we
  have $p_{\mathrm{dark}} \simeq 2 p_d$).}.

For the SARG protocol, taking into account the two-photon
contribution in the lower bound allows to increase the lower bound.
In the case of SARG without decoy states, we could thus improve
significantly the bound of \cite{FuTaLo}. Our conclusion is
therefore different: we find that the SARG protocol performs better
than BB84 for high visibility $V \simeq 1$ (see Fig.
\ref{fig_no_decoy_V_1}). However, the SARG is more sensitive to the
loss of the channel, and for $V = 0.95$ for instance, BB84 is more
efficient (Fig. \ref{fig_no_decoy_V_095}).

In the case of SARG with decoy states, the two-photon contribution
had already been taken into account in \cite{FuTaLo}, and we again
get similar results. However, we could slightly improve the rate
with the improved calculation of $S_1^{SARG}(Q_1)$ (see Appendix
C), and with the local randomization process. Nevertheless, our
conclusion is the same as in \cite{FuTaLo}, namely that when decoy
states are used, the SARG is outperformed by the BB84 protocol.


\section{Further applications, and open problems}

There are still several possibilities to improve the lower bounds on
the secret-key rate of QKD protocols. One way to look at this
problem is to analyze the properties of the set $\Gamma$ over which
one has to optimize in order to obtain the lower bound (see e.g.
Eq.~\eqref{rategen}). Concerning the single photon QKD protocols,
one might try to find the conditions on the encoding (and decoding)
operations which would lead to a properly restricted set $\Gamma_Q$,
such that a high QBER can be tolerated.

In a protocol based on weak coherent pulses, it might be advantageous
to take the detected double clicks into account. As mentioned above,
this would (most likely) impose further restrictions on the set of
possible attacks and thus result in an improvement of the secret-key
rate. In addition, it would be interesting to generalize the ideas
developed in this article to a scenario, where not only the intensity
of light is used but where also the coherence of the light is checked
(similar to the decoy states). One protocol taking this into account
has for instance been proposed in~\cite{COW}.  Another possibility is
to consider protocols based on weak coherent pulses that use two-way
post-processing, as studied by Lo~\cite{Lo06}. We also note here that
the techniques presented here can also be applied to protocols based
on squeezed states.

In this work, we considered the so-called \emph{untrusted-device
  scenario}, where the adversary might arbitrarily modify the
efficiency of Bob's detector. If one considers the reasonable
situation, where Eve cannot influence Bob's device, one might
obtain larger values for the key rate.

\bigskip

\section{Acknowledgements}

The authors would like to thank Nicolas Gisin, Antonio Acin, and
Valerio Scarani for helpful discussions. This project is partly
supported by SECOQC and by the FWF. RR acknowledges support by HP
Labs, Bristol and BK by the FWF through the Elise-Richter project.

\appendix

\section{Proof of Lemma~\ref{lem:prod}} \label{app:proof}

In this appendix we prove the Lemma presented in Section III. The
operator $\sigma^{\otimes n}$ is symmetric and can thus be written as
$\sigma^{\otimes n} = \sum_{n'_1, \ldots, n'_4} \mus_{n'_1, n'_2,
  n'_3, n'_4} {\cal P}_S (P_{\ket{\Phi_{00}}}^{\otimes n'_1}\otimes
P_{\ket{\Phi_{01}}}^{\otimes n'_2}\otimes P_{\ket{\Phi_{10}}}^{\otimes
  n'_3}\otimes P_{\ket{\Phi_{11}}}^{\otimes n'_4})$, for appropriate
coefficients $\mus_{n'_1,n'_2,n'_3,n'_4}$.  Hence, with the definition
$p := \mus_{n_1, n_2, n_3, n_4}$, we have
  \[
    \sigma^{\otimes n}
  =
    p \rho_n + (1-p) \tilde{\rho}_n
  \]
  where $\tilde{\rho}_n$ is a symmetric quantum state on $n$
  subsystems. Moreover, it is easy to see that the coefficient $p$
  cannot be smaller than $\frac{1}{n}$.

  By linearity, we get the following expression for the state after
  the operation $\cE^{\otimes \bn}$ has been applied to
  $\sigma^{\otimes n}$:
  \begin{equation} \label{eq:afteroperation}
    \bar{\sigma}^{\otimes \bn}
  =
    \cE^{\otimes \bn}(\sigma^{\otimes n})
  =
    p \cE^{\otimes \bn}(\rho_n) + (1-p) \cE^{\otimes \bn}(\tilde{\rho}_n) \ .
  \end{equation}

  Because $\bar{\sigma}^{\otimes \bn}$is symmetric, it can be written
  as $ \bar{\sigma}^{\otimes \bn} = \sum_{\bn_1, \ldots, \bn_4}
  \mub'_{\bn_1,\bn_2,\bn_3,\bn_4} {\cal P}_S
  (P_{\ket{\Phi_{00}}}^{\otimes \bn_1}\otimes
  P_{\ket{\Phi_{01}}}^{\otimes \bn_2}\otimes
  P_{\ket{\Phi_{10}}}^{\otimes \bn_3}\otimes
  P_{\ket{\Phi_{11}}}^{\otimes \bn_4})$, for some coefficients
  $\mub'_{\bn_1,\bn_2,\bn_3,\bn_4}$.  Furthermore, by the law of large
  numbers, the sum of the coefficients
  $\mub'_{\bn_1,\bn_2,\bn_3,\bn_4}$ for tuples
  ${\bn_1,\bn_2,\bn_3,\bn_4}$ which are not contained in
  $\mathcal{B}^\eps(\lambdab_1, \ldots, \lambdab_4)$ is exponentially
  small, i.e.,
  \begin{equation} \label{eq:chernoff}
    \sum_{(\bn_1, \ldots, \bn_4) \notin \mathcal{B}^\eps(\lambdab_1, \ldots,
\lambdab_4)} \mub'_{\bn_1,\bn_2,\bn_3,\bn_4}
  \leq
    2^{-\Theta(\bn \eps^2)} \ .
  \end{equation}
  Finally, because of~\eqref{eq:afteroperation},
  \[
     \mub'_{\bn_1,\bn_2,\bn_3,\bn_4} \geq p \cdot \mub_{\bn_1,\bn_2,\bn_3,\bn_4} \
,
  \]
  where $\mub_{\bn_1,\bn_2,\bn_3,\bn_4}$ are the coefficients of
  $\rhob_n$. Since $p \geq \frac{1}{n}$,
  \[
    \mub_{\bn_1,\bn_2,\bn_3,\bn_4} \leq n \cdot \mub'_{\bn_1,\bn_2,\bn_3,\bn_4} \
.
  \]
  Combining this with~\eqref{eq:chernoff}, we conclude
  \[
 \sum_{(\bn_1, \ldots, \bn_4) \notin \mathcal{B}^\eps(\lambdab_1, \ldots,
\lambdab_4)} \mub_{\bn_1,\bn_2,\bn_3,\bn_4}
  \leq
    n 2^{-\Theta(\bn \eps^2)} \ .
  \]
\qed

\section{Advantage distillation using the XOR process} \label{app:xor}

In this appendix we explain how the XOR process applied to many
qubit pairs can be easily included within this formalism. Alice
selects randomly a set of bits and informs Bob about this set.
Then, Alice and Bob compute both the XOR of those bits and keep
only the result, discarding all the others. Our goal is to find a
simple description of the remaining logical bits, Eve's system,
and the classical information sent form Alice to Bob (note that
Eve knows the randomly chosen set which is used by Alice and Bob).
We demonstrate here how this can be achieved with the example of
three qubit pairs. The idea can be easily generalized to any
number of pairs.

Quantum mechanically the XOR operation can be described by a
controlled-not operation, denoted by $Uc$. Three copies of the state
$\ket{\Psi}_{ABE}=\sum_{i,j}\sqrt{\lambda_{i,j}}\ket{\Phi_{i,j}}_{AB}\ket{\Phi_{i,j}}_E$
transform, under the transformation $Uc_{A}^{3\rightarrow
  1}Uc_{A}^{2\rightarrow 1}\otimes Uc_{B}^{3\rightarrow
  1}Uc_{B}^{2\rightarrow 1}$ to the state \bea
\sum_{i,j,k,l,m,n}\sqrt{\lambda_{i,j}\lambda_{k,l}\lambda_{m,n}}
\ket{\Phi_{i+k+m,j}}_{A_1B_1}\\
\nonumber
\ket{\Phi_{k,l+j}}_{A_2B_2}\ket{\Phi_{m,n+j}}_{A_3B_3}\ket{\Chi_{i,j,k,l,m,n}}_E,\eea
where
$\ket{\Chi_{i,j,k,l,m,n}}_E=\ket{\Phi_{i,j}}\ket{\Phi_{k,l}}\ket{\Phi_{m,n}}$.
Since Alice and Bob are not going to use the systems $2$ and $3$
anymore, we want to consider a state that describes only Alice's and
Bob's first systems. More importantly, we want to give Eve a
purification of this state. If we would assume that Eve has a
purification of the state describing systems $A_1$ and $B_1$, this
would be equivalent to assume that Eve has Alice' and Bob's second and
third pair after this transformation. It is evident that we assume
then that she has more power than she actually has. In order to avoid
to give her too much power we use the idea mentioned in Section
\ref{improve_bound} (see also \cite{KrRe05}), by considering the
systems $A_2,B_2,A_3,B_3$ as auxiliary system $R$ \footnote{A similar
  argument has also been used in \cite{GoLo03,Ch02} to include this
  process.}. For the unitary transformations, $U_{k,l,m,n}$ we choose
$U_{i,j,k,l,m,n}=\sigma^{A_1}_z$ for $l+j=n+j=1$ and the identity
otherwise. It can be easily verified that the state describing Alice's
and Bob's first system is then the partial trace over $E,R$ of the
state $\ket{\tilde{\Psi}}_{A_1B_1 R E}= \sum_{i,j,k,l,m,n}
\sqrt{\lambda_{ij}\lambda_{kl}\lambda_{mn}}$$\ket{\Phi_{i+k+m,j+\delta_{l+j,1}\delta_{n+j,1}}}_{A_1B_1}
$ $\ket{\phi_{j,k,l,m,n}}_R\ket{\Chi_{i,j,k,l,m,n}}_E,$ where
$\ket{\phi_{j,k,l,m,n}}_R$ denotes the state
$\ket{\Phi_{k,l+j}}_{A_2B_2}\ket{\Phi_{m,n+j}}_{A_3B_3}$. As explained
in Section \ref{improve_bound}, providing Eve with a purification of
the state that describe the systems $A_1,B_1$ never underestimates her
power. The eigenvalues of the two-qubit Bell-diagonal state describing
Alice's and Bob's remaining systems, denoted by
$\tilde{\lambda}_{i,j}$ are \bea
\tilde{\lambda}_{i,j}&=&\lambda_{i,j}^2(\lambda_{i,j}+3\lambda_{i,j+1})+3\lambda_{i+1,j}^2(\lambda_{i,j}+\lambda_{i,j+1})\\
\nonumber &+&6\lambda_{i,j}\lambda_{i+1,j}\lambda_{i+1,j+1}.\eea

The intuition for this choice of unitary transformations is the
following. The state $\ket{\Psi}_{ABE}$ under consideration is
supposed to lead to a secret-bit. Thus, the coefficients
$\lambda_{i,j}$ are such that it is very likely that if both,
$l+j=1$ and $n+j=1$ then $j=1$, which means that within the
remaining qubit-pair there is a phase-flip error. The unitaries
are chosen such that this error is corrected.

Using the new eigenvalues of the state describing Alice' and Bob's
remaining bits, it is straightforward to compute the lower bound on
the secret-key rate (Eq.~\eqref{rategen}).

\section{An improved analysis of the SARG protocol with single photons} \label{app:SARGimprove}

In the SARG protocol the bit value $0$ ($1$) is encoded in the
$z$-basis ($x$-basis) respectively. During the sifting phase Alice
announces a set containing two states, the one which she sent and
one in the other basis. There are $4$ different encoding and
decoding operators. For instance
$A_1=\ket{0}\bra{0_z}+\ket{1}\bra{0_x}$ and
$B_1=\ket{0}\bra{1_x}+\ket{1}\bra{1_z}$ describe the situation where
Alice sends on of the two states $\{\ket{0_z},\ket{0_x}\}$ and tells
Bob that the sent state is within this set. Let us for the moment
consider a single qubit sent by Alice (for more details see
\cite{BrGi05}). The state shared by Alice, Bob, and Eve after the
sifting is given by $\ket{\chi}_{ABER_1}= \sum_j A_j\otimes B_j
\ket{\Psi}_{ABE}\ket{j}_{R_1}$, where $\ket{\Psi}_{ABE}$ is the
state shared by Alice, Bob, and Eve after Eve´s intervention. Now,
we apply some symmetrization to the state, which does not change any
security consideration, as explained in Section II. Let us consider
the state
$\ket{\tilde{\chi}}_{ABER_1R_2}=\ket{\chi}_{ABER_1}\ket{0}_{R_2}+\sigma_z^A\otimes
\sigma_z^B \ket{\chi}_{ABER_1}\ket{1}_{R_2}$. It is straightforward
to show that the reduced state describing Alice' and Bob's system is
equal to $ \tilde{\cD}_2 {\cD_1 [\cD_2(\rho_0)]}$, with
$\rho_0=\tr_E(P_{\Psi})$. Here, $\tilde{\cD}_2(\rho)=
1/2(\rho+\sigma_z\otimes \sigma_z \rho \sigma_z\otimes \sigma_z)$,
$\cD_1(\rho)=\sum_j A_j\otimes B_j \rho A_j^\dagger\otimes
B_j^\dagger$ is given by the protocol and $\cD_2$ denotes the
depolarizing map, i.e. $\cD_2(\rho)=1/4(\rho+\sigma_x\otimes
\sigma_x \rho \sigma_x\otimes \sigma_x+\sigma_y\otimes \sigma_y \rho
\sigma_y\otimes \sigma_y+\sigma_z\otimes \sigma_z \rho
\sigma_z\otimes \sigma_z)$. Furthermore, the action of $\cD_1$ on a
Bell-diagonal state is the same as $A_1\otimes B_1$ on that state.
Thus, we only need to consider the situation where Eve has a
purification of the state $\cD_2(\rho_0)$, i.e. the state before the
action of $\cD_1$ and $\tilde{\cD}_2$. Using the results of
\cite{KrRe05,ReKr05} this implies that the state we have to use in
order to compute the lower bound on the secret-key rate is
$\rho_{ABE}=\tilde{\cD}_2^{AB} (P_{A_1\otimes
B_1\ket{\Phi}_{ABE}})$, where
$\ket{\Phi}_{ABE}=\sqrt{\lambda_{00}}\ket{\Phi_{00}}_{AB}\ket{\Phi_{00}}_E
+\sqrt{\lambda_{01}}\ket{\Phi_{01}}_{AB}\ket{\Phi_{01}}_E+
\sqrt{\lambda_{10}}\ket{\Phi_{10}}_{AB}\ket{\Phi_{10}}_E+\sqrt{\lambda_{11}}\ket{\Phi_{11}}_{AB}\ket{\Phi_{11}}_E$,
i.e a purification of the Bell-diagonal state $\cD_2(\rho_0)$.

Using this description it is straightforward to compute the state
describing Alice' and Bob's system, which is, in contrast to former
considerations, no longer Bell-diagonal. In the following we
consider the situation where Bob accepts only if the probability for
him to obtain the bit values $0$ is the same as detecting $1$. This
is a first step in the parameter estimation. Note that this
condition imposes $\lambda_{01}=\lambda_{10}$. The QBER, $Q$, can be
easily determined and one finds
$Q=(\lambda_{01}+\lambda_{11})/(1/2+\lambda_{01}+\lambda_{11})$.
Using the normalization condition we find that the coefficients in
the state $\ket{\Phi}_{ABE}$ are given by:
$\lambda_{00}=1-Q/(1-Q)+\lambda_{11},
\lambda_{01}=Q/(2(1-Q))-\lambda_{11}, \lambda_{10}=\lambda_{01}$.
Thus, for a fixed QBER there is only one parameter, $\lambda_{11}
\in [0,Q/(2(1-Q))]$, over which one needs to minimize to obtain the
lower bound on the secret-key rate given in formula
Eq.~\eqref{rategenmixed}. Without the local randomization one finds
that the lower bound on the secret-key rate is positive as long as
$Q\leq 0.1167$. Including the local randomization allows to increase
the tolerable QBER to $0.1308$ compared to the previously known
bounds of $0.0968$ without and $0.1095$ with local randomization,
respectively \cite{BrGi05}.

\section{Calculations related to the analysis of protocols based on coherent pulses}

\label{app_comput_LB_wcp}

\label{app_expected_rates}

This appendix contains some calculations related to the evaluation
of the lower bound~\eqref{rateweak} on the secret-key rate for the
BB84 and SARG protocols with weak coherent pulses (see
Section~\ref{sec:coh}).

For this purpose, we first compute the infimum $S_n(Q_n) :=
\inf_{\sigma_n \in \Gamma_{Q_n}} S(X|E,n)$ for any given $Q_n$, and
then optimize (from Eve's point of view) over the parameters $R_n,
Q_n$. These parameters must be compatible with the measurable
quantities $R_\mu, Q_\mu$: in the case of protocols which do not use
decoy states, this leads to particular constraints for each
protocol, which we derive here. (Note that for protocols with decoy
states, Alice and Bob can estimate all rates $R_n, Q_n$: Eve can no
longer optimize over these parameters.)

Recall that we work in the untrusted device scenario, where Eve has
full control over Bob's detectors. Dark counts do not occur, and
therefore $R_0 = 0$, as Eve should obviously not send any photon to
Bob when she receives an empty pulse from Alice. Moreover, we
consider protocols where Bob treats all double clicks as if only one
randomly chosen detector clicked.

In a second step, in order to give estimations of our bounds, we
compute the typical values of the yields and error rates if no
adversary is present, i.e., if the channel between Alice and Bob is
a depolarizing channel with fidelity F (or disturbance $D = 1 - F$)
and with a transmission factor $t$. In addition, we suppose in that
case that Bob's detectors have an efficiency $\eta_{det}$ and a
probability of dark counts $p_d$. We will use the notations $\eta =
t \eta_{det}$ for the overall transmission factor and $\bar{p}_d = 1
- p_d$.
%

\subsection{BB84 protocol}

\subsubsection{Eve's uncertainty on the one-photon pulses}

For BB84, the set $\Gamma_{Q_1}$ contains all states with diagonal
entries (in the Bell basis) $\lambda_{00} = 1 - 2Q_1 + \lambda_{11}$
and $\lambda_{01} = \lambda_{10} = Q_1 - \lambda_{11}$, for any
$\lambda_{11} \in [0,Q_1]$ \cite{KrRe05,ReKr05}.

One can easily prove that $S(X|E,n=1)$ takes its minimum when
$\lambda_{1,1} = Q_1^2$. Then, a straightforward calculation shows
that $S_1^{BB84}(Q_1) = \inf_{\sigma_1 \in
  \Gamma_{Q_1}} S(X|E,n=1) = 1 - h(Q_1)$. Note that
  $S_1^{BB84}(Q_1)$ is decreasing for $0 \leq Q_1 \leq 1/2$: as expected, the
  higher the error Eve introduces, the more she reduces her
  uncertainty.

\subsubsection{Constraints on the yields and error rates}

In the BB84 protocol, the probability that Alice and Bob choose the
same basis for their preparation and measurement respectively is
$1/2$ (this is the sifting factor). Therefore we have $Y_n \leq
\frac{1}{2}$ for all $n$, which implies the following bounds:

\bea R_1 & = & p_1 Y_1 \leq \frac{1}{2} p_1 \\ R_1 & = & R_{\mu} -
\sum_{n \geq 2} p_n Y_n \geq R_{\mu} - \frac{1}{2} \sum_{n \geq 2}
p_n. \eea These are the first two constraints announced in
\eqref{constr_BB84}. The third constraint follows from the
definition of $Q_{\mu}$, $R_{\mu} Q_{\mu} = \sum_{n}R_n Q_n$.

\subsubsection{Yields and error rates for depolarizing channels}

When implementing the BB84 protocol, Alice and Bob would estimate
the quantities $Q_{\mu}$, $R_{\mu}$ and then compute the rate as
explained above. In order to get an idea how good the obtained
bounds on the rate are we evaluate here these quantities for the
situation where there is no Eve present and Alice and Bob share a
lossy depolarizing channel.

In BB84, when Alice sends $n$ photons, the probability that Bob
chooses the same basis as Alice and gets a single or a double click is
: \[
\begin{array}{lll} Y_n & = & \frac{1}{2} \big[ 1 - \bar{p}_d^{\ 2}
(1-\eta)^n \big]
\end{array}
 \]

Bob gets a wrong bit if only the wrong detector clicks, or if the
two detectors click, but he randomly chooses a wrong bit. This
happens with probability :

\[
\begin{array}{lll} Y_n Q_n & = & \frac{1}{2} \sum_{k=0}^n \mathrm{C}_n^k F^k
D^{n-k}
\Big[  [\bar{p}_d(1-\eta)^k] [1-\bar{p}_d(1-\eta)^{n-k}] \\
& & \qquad \qquad + \frac{1}{2} [1 - \bar{p}_d(1-\eta)^k] [1-\bar{p}_d(1-\eta)^{n-k}] \Big] \\
& = & \frac{1}{4} \big[ 1 + \bar{p}_d (1-F \eta)^n - \bar{p}_d (1-D
\eta)^n - \bar{p}_d^{\ 2} (1-\eta)^n \big]
\end{array}
 \]
 
 When Alice uses a Poissonian source (i.e.\ $p_n =
 \frac{\mu^n}{n!}e^{-\mu}$), the overall yield and error rate are then
\[\begin{array}{lll}
R_{\mu} & = & \frac{1}{2} \big[ 1 - \bar{p}_d^{\ 2} \ e^{-\mu \eta} \big] \\
R_{\mu} Q_{\mu} & = &  \frac{1}{4} \big[ 1 + \bar{p}_d e^{-\mu F
\eta} - \bar{p}_d e^{-\mu D \eta} - \bar{p}_d^{\ 2} \ e^{-\mu
\eta} \big].
\end{array}
\]

\subsection{SARG protocol}

\subsubsection{Eve's uncertainty on the one-photon pulses}

In order to compute Eve's uncertainty on the one-photon pulses, we
use the method presented in Appendix C. We don't have an
analytical expression for $S_1^{SARG}(Q_1) = \inf_{\sigma_1 \in
  \Gamma_{Q_1}} S(X|E,n=1)$, but we compute it numerically. Note that we find $S_1^{SARG}(Q_1)$
  is decreasing only for $0 \leq Q_1 \lesssim 0.338$, and does not reach zero.

\subsubsection{Eve's uncertainty on the two-photon pulses}


We follow the calculations of~\cite{FuTaLo} to compute Eve's
uncertainty on the two-photon pulses. The set $\Gamma_{Q_2}$ contains
all states with the following diagonal entries (in the Bell basis)
\bea
\begin{array}{l}
\lambda_{00} = 1 - Q_2 - \lambda_{01} \\
\lambda_{10} = Q_2 - \lambda_{11} \\
\lambda_{01} + \lambda_{11} \leq x Q_2 + g(x), \forall x
\end{array} \eea
where $g(x) = \frac{1}{6}(3 - 2x + \sqrt{6 - 6 \sqrt{2}x + 4x^2})$
\cite{FuTaLo}. When minimizing $x Q_2 + g(x)$ over $x$, we get

\bea
\begin{array}{l}
\lambda_{00} = 1 - Q_2 - \lambda_{01} \\
\lambda_{10} = Q_2 - \lambda_{11} \\
\lambda_{01} + \lambda_{11} \leq B(Q_2)
\end{array} \eea
where $B(Q_2) = \frac{1}{2} +
\frac{1}{2}\sqrt{Q_2(1-\frac{3Q_2}{2})} -
\frac{\sqrt{2}}{4}(1-3Q_2)$.

One can show that for $Q_2 \leq \frac{1}{6}$, $B(Q_2) \leq
\frac{1}{2}$ and the optimal choice of the parameters $\lambda_{ij}$
for Eve is $\lambda_{01} + \lambda_{11} = B(Q_2)$ (i.e. Eve should
make the phase error as high as possible, up to $\frac{1}{2}$), and
$\lambda_{11} = Q_2 B(Q_2)$. Then, a straightforward calculation
gives $S_2^{SARG}(Q_2) = \inf_{\sigma_2 \in
  \Gamma_{Q_2}} S(X|E,n=2) = 1 - h(B(Q_2))$.
Note that $S_2^{SARG}(Q_2)$ is decreasing for $0 \leq Q_2 \leq
\frac{1}{6}$, and $S_2^{SARG}(\frac{1}{6}) = 0$.

\subsubsection{Constraints on the yields and error rates}

In the case of SARG, because of the non orthogonality of the quantum
states that are used to encode the classical bit values, it is a
little bit more tricky to find the constraints that the yields and
error rates must satisfy. Here, we will derive a constraint on the
yields without errors (or probability that Bob gets a right
conclusive result), i.e., on $p_{\mathrm{right}} = Y_n(1-Q_n)$ (for
any $n \in \bbN$).

To this aim, let's suppose in a first step that Alice sends photons in
the state $\ket{+z}$, that Eve attacks the pulse and decides either to
forward one photon to Bob in the state $\rho_B$, or to block the
pulse. In this case, Bob gets a right conclusive result if (i) Alice
announces the set $\{\ket{+z}, \ket{+x}\}$ (which she does with
probability 1/2), Bob chooses to measure $\sigma_x$ (probability 1/2)
and (only) the detector corresponding to $\ket{-x}$ clicks ; or (ii)
Alice announces the set $\{\ket{+z}, \ket{-x}\}$, Bob chooses to
measure $\sigma_x$, and the detector corresponding to $\ket{+x}$
clicks. Therefore, Bob's probability to get a right conclusive result
when Alice sends $\ket{+z}$ is bounded by: \bea p_{\mathrm{right}|+z}
& \leq & \frac{1}{4} \bra{-x} \rho_B \ket{-x} + \frac{1}{4} \bra{+x}
\rho_B \ket{+x} \\ & \leq & \frac{1}{4} \ Tr(\rho_B) = \frac{1}{4}.
\eea

This result actually does not depend on the state sent by Alice, and
we therefore have \bea p_{\mathrm{right}} = Y_n(1-Q_n) \leq
\frac{1}{4}. \eea

\bigskip

The first three constraints announced in~\eqref{constr_sarg} then
follow :

\bea R_1 (1 - Q_1) & \leq & \frac{1}{4} p_1 \\
R_2 (1 - Q_2) & \leq & \frac{1}{4} p_2 \\
R_1 (1 - Q_1) + R_2 (1 - Q_2) & \geq & R_{\mu} (1 - Q_{\mu}) -
 \\ & & \  - \frac{1}{4} \sum_{n \geq 3} p_n
\eea

As before, the last constraint follows from the definition of
$Q_{\mu}$.

\subsubsection{Yields and error rates for depolarizing channels}

As for the BB84--protocol, we evaluate here the lower bound on the
secret key rate for the situation where there is no Eve present
and Alice and Bob share a lossy depolarizing channel, in order to
get an idea of how good the obtained bounds on the rate are.

In order to calculate the yields and error rates for the SARG
protocol, let's suppose that Alice sends $n$ photons in the state
$\ket{+z}$, and announces $\{ \ket{+z},\ket{+x} \}$. By symmetry, the
following still holds for any state sent by Alice, and any
announcement. Similar calculations can be found in \cite{BrGi05}.

If Bob measures $\sigma_z$, he gets a (wrong) conclusive click on
the detector corresponding to $\ket{-z}$, or a double click with
probabilities:

\[\begin{array}{lll}
p_{\ket{-z}|z} & = & \sum_{k=0}^n \mathrm{C}_n^k F^k D^{n-k}
[\bar{p}_d(1-\eta)^k] [1-\bar{p}_d(1-\eta)^{n-k}] \\
& = & \bar{p}_d (1-F \eta)^n - \bar{p}_d^{\ 2} (1-\eta)^n \\
p_{2 clicks|z} & = & 1 - \bar{p}_d (1-F \eta)^n - \bar{p}_d (1-D
\eta)^n + \bar{p}_d^{\ 2} (1-\eta)^n
\end{array}
 \]

Similarly, if Bob now measures $\sigma_x$, he gets a (right)
conclusive click on the detector corresponding to $\ket{-x}$, or a
double click with probabilities:

\[\begin{array}{lll}
p_{\ket{-x}|x} & = & \bar{p}_d (1-\frac{\eta}{2})^n - \bar{p}_d^{\
2} (1-\eta)^n \\
p_{2 clicks|x} & = & 1 - 2 \bar{p}_d (1-\frac{\eta}{2})^n +
\bar{p}_d^{\ 2} (1-\eta)^n
\end{array}
 \]

Since Bob randomly chooses the basis he measures, with equal
probabilities, and since he randomly chooses one outcome in the case
of double clicks (conclusive or not), then the probability that
Bob's result is conclusive when Alice sends $n$ photons is $Y_n =
\frac{1}{2} \big(p_{\ket{-z}|z} + \frac{1}{2} p_{2 clicks|z}\big) +
\frac{1}{2} \big(p_{\ket{-x}|x} + \frac{1}{2} p_{2 clicks|x}\big)$,
and the error rate on these pulses is $Y_n Q_n = \frac{1}{2}
\big(p_{\ket{-z}|z} + \frac{1}{2} p_{2 clicks|z}\big)$. We find:

\[\begin{array}{lll}
Y_n & = & \frac{1}{2} \big[ 1 + \frac{\bar{p}_d}{2}
(1-F \eta)^n - \frac{\bar{p}_d}{2} (1-D \eta)^n - \bar{p}_d^{\ 2} (1-\eta)^n \big]\\
Y_n Q_n & = & \frac{1}{4} \big[ 1 + \bar{p}_d (1-F \eta)^n -
\bar{p}_d (1-D \eta)^n - \bar{p}_d^{\ 2} (1-\eta)^n \big]
\end{array}
 \]

 For a Poissonian source, the overall yield and error rate are then
\[\begin{array}{lll}
R_{\mu} & = & \frac{1}{2} \big[ 1 + \frac{\bar{p}_d}{2}
e^{-\mu F \eta} - \frac{\bar{p}_d}{2} e^{-\mu D \eta} - \bar{p}_d^{\ 2} \ e^{-\mu \eta} \big] \\
R_{\mu} Q_{\mu} & = & \frac{1}{4} \big[ 1 + \bar{p}_d e^{-\mu F
\eta} - \bar{p}_d e^{-\mu D \eta} - \bar{p}_d^{\ 2} \ e^{-\mu \eta}
\big].
\end{array}
\]

\end{document}